\documentclass[letterpaper,9pt,twocolumn]{autart}
\usepackage[]{graphicx}
\usepackage{cite}
\usepackage[dvips]{color}
\usepackage{amsfonts}
\usepackage{amsmath}
\usepackage{amssymb}
\usepackage{cite}
\usepackage{subfigure}
\usepackage{psfrag}
\usepackage[thinlines]{easymat}
\usepackage[dcucite]{harvard}

\newtheorem{theorem}{\textbf{Theorem}} 

\newtheorem{corollary}[theorem]{\textbf{Corollary}}
\newtheorem{definition}[theorem]{\textbf{Definition}}

\newtheorem{assumption}[theorem]{\textbf{Assumption}}
\newtheorem{problem}[theorem]{\textbf{Problem}}
\newtheorem{remark}[theorem]{\textbf{Remark}}

\newcommand{\subscr}[2]{#1_{\textup{#2}}} 
\newcommand{\subupscr}[3]{#1_{\textup{#2}}^{\textup{#3}}} 
\begin{document}
\runauthor{}
\begin{frontmatter}
\title{Visibility\hspace{0.22cm}maintenance\hspace{0.22cm}via\hspace{0.22cm}controlled\hspace{0.22cm}invariance\vspace{0.05cm}\\
for\hspace{0.22cm}leader-follower\hspace{0.22cm}Dubins-like\hspace{0.22cm}vehicles}


\thanks[footnoteinfo]{Corresponding author. 
This material is based upon work supported in part by ONR Award N00014-07-1-0721.
This \mbox{paper} was not presented at any IFAC meeting.}

\author[UTA]{Fabio Morbidi$^{\star}$}\ead{morbidi@dii.unisi.it},
\author[UCSB]{Francesco~Bullo}\ead{bullo@engineering.ucsb.edu},
\author[Siena]{Domenico~Prattichizzo}\ead{prattichizzo@dii.unisi.it}
\address[UTA]{Dept. of Computer Science and Engineering,
\mbox{University of Texas at Arlington}, TX, 76019, USA}
\address[UCSB]{Dept. of Mechanical Engineering and Center for Control, Dynamical Systems
and Computation, University of California at Santa Barbara, CA, 93106-5070, USA.}
\address[Siena]{Dept. of Information Engineering, University of Siena,
  53100 Siena, Italy}

\begin{keyword} 
Autonomous mobile robots; Visibility maintenance; Leader-follower; Controlled invariance; Input constraints  
\end{keyword}

\begin{abstract}
The paper studies the \emph{visibility maintenance problem} (VMP) for
a leader-follower pair of Dubins-like vehicles with input constraints, and
proposes an original solution based on the notion of controlled
invariance. The nonlinear model describing the relative dynamics of the vehicles
is \mbox{interpreted} as linear uncertain system, with the leader robot acting 
as an external disturbance. The VMP is then reformulated as a \emph{linear constrained
regulation problem with \mbox{additive} disturbances} (DLCRP).
Positive \mbox{$\mathcal{D}$-invariance} conditions for
linear uncertain systems with parametric \mbox{disturbance} matrix
are introduced and used to solve the VMP when box
bounds on the state, control input and disturbance are considered.
The~proposed design procedure is shown to be easily adaptable 
to more general working scenarios. 
Extensive simulation results are provided to illustrate the theory 
and show the effectiveness of our approach.
\end{abstract}
\end{frontmatter}
\thispagestyle{empty} 
\pagestyle{empty} 
\pagestyle{plain}

\vspace{-0.2cm}
\section{Introduction}
\label{SEC:intro}\vspace{-0.1cm}

\subsection{Problem description and motivation}
\vspace{-0.25cm}

This paper considers a number of visibility maintenance problems between 
autonomous vehicles. The~simplest formulation is a leader-follower setup, 
in which both leader and follower are nonholonomic vehicles constrained 
to move along planar paths of bounded curvature, with limited positive
forward speed. 
The follower vehicle's goal is to maintain the leader inside an appropriate 
sensing region.
The theory of controlled invariance for uncertain linear systems
is shown to be well suited for this objective as well as for achieving more involved tasks, such as, e.g.,
simultaneously reject unknown but bounded disturbances or 
preserve visibility in multi-vehicle chain formations.\\
We have two main motivations for the visibility maintenance problems studied in this article. 
First, we are interested in surveillance and patrolling problems with
formations of robotic vehicles 
in aerial and ground environments. We~envision scenarios where a robot equipped with
sensors with limited sensing footprints (such as, e.g., panoramic cameras, laser
range finders, or high resolution radars) moves in such a way as to
maintain a second moving target within its field of view. 
Second, this work is motivated by the need to design network-wide visibility 
and connectivity maintenance algorithms for groups of robotic vehicles. 
In~the multi-agent network domain, connectivity is indeed a~classic requirement 
necessary to guarantee the correct completion of numerous distributed algorithms.

\vspace{-0.05cm}
\subsection{Literature review}
\vspace{-0.15cm}

In the context of \emph{visibility maintenance} between pairs of vehicles, two distinct
literature domains are relevant to this work. First, visibility-based
pursuit-evasion problems for robots in complex environments have been
investigated in continuous-time
in~\cite{GuibasLaLaLiMo_LNCS97,GerkeyThGo_IJRR06,BhattacharyaHu_IJRR10} 
and in discrete-time in~\cite{IslerKaKh_TRO05}: 
in~these works the vehicles' dynamic models are elementary and the 
proposed solutions are not applicable to nonholonomic vehicles with
limited positive forward speed. 
Second, the vast literature on aircraft pursuit-evasion has focused much attention to
game theory, optimal control, and numerical algorithms. 
A~hybrid-systems and game-theoretic approach to
aircraft conflict resolution is pursued in~\cite{TomlinLySa_IEEE00}.
Differential game problems between aircraft are discussed
in~\cite{MerzHa77,JarmarkHi_JGCD84,ShimaSh_JGCD02}. 
In~\cite{Glizer99}, a planar pursuit-evasion problem in which the
target set is defined by a capture radius and constraints on the angular state variables 
(line-of-sight angle) is analyzed, and a necessary and sufficient condition for capture of the evader 
from any initial state is established using a variational method. However,
differently from the problem studied in this paper, the author
considers constant positive forward speed for 
the nonholonomic vehicles and unlimited turning rate for the evader.
More recently, in~\cite{MazoSpJoHu_ICRA04},
the problem of estimating and tracking the motion
of a moving target by a team of unicycles equipped with directional sensors with
limited range, is addressed using a hierarchical control scheme.\\ 
In the context of \emph{connectivity maintenance} in multi-agent networks, the
literature has experienced a recent spurt of growth (we refer
to~\cite{Book_BulloCoMa08,book_Egerstedt10} 
for recent surveys on this topic): 
two typical multi-agent tasks requiring network connectivity are ``consensus''
\cite{Moreau_TAC05,OlfatiFaMu_IEEE07} and ``rendezvous''
\cite{CortesMaBu_TAC06,LinMoAn_SICON07}.
In this active research area, robots with limited communication capability
are often modeled as transmitters with disks of finite radius. As~the
vehicles move to achieve a goal, it is generally hard to guarantee
the connectivity among the members of the group is preserved over time. 
In~terms of design, it is then required to constrain robots' control inputs such that 
the resulting topology is always connected throughout its course of evolution. 
\emph{Potential fields} and \emph{geometric optimization methods} are the standard tools used in the 
literature to address the connectivity maintenance problem: 
a~list of key references, yet far from being complete, is
\cite{AndoOaSuYa_TRA99,SpanosMu_ACC05,JiEg_TRO07,ZavlanosPa_TRO07,DimarogonasKy_TRO08,SavlaNoBu_JCO09,YangFrGoLySrSu_AUTO10}. 
These works differ from the setup proposed in this paper
in at least two important ways: first, the vehicle's dynamics is assumed to
be locally controllable, and second, connections between two robots are
bidirectional. In our problem, instead, we consider nonholonomic vehicles that are
not locally controllable (they move forward with positive speed, along paths
of bounded curvature) and we deal with sensor footprints such that the
visibility links are not bidirectional. \mbox{We finally} point out that most 
of the works cited above do not explicitly account for robots' input
constraints. 
\subsection{Original contributions}
\vspace{-0.25cm}

The basic setup considered in this paper consists of two nonholonomic agents: 
a \emph{leader} (or evader) $\textup{L}$ and a \emph{follower} (or pursuer)
$\textup{F}$.  The robots can rotate with bounded angular velocity, but
\mbox{similarly} to Dubins' vehicles~\cite{Dubins57} can only move forward.
The follower is equipped with a sensing device characterized by a
\emph{visibility set}~$\mathcal{S}$, a compact and convex polyhedral region
encoding both the position and angle information.  The leader moves along
a given trajectory and the follower aims at maintaining $\textup{L}$
always inside its visibility set~$\mathcal{S}$,
while respecting suitable bounds on the control~inputs.
Inspired by~\cite{TiwariFuBhMu_CDC04}, where the concept of cone invariance
is used to solve the multiagent rendezvous problem and by the results
in~\cite{Blanchini_TAC90,Blanchini_JOTA91}, this paper addresses the
\emph{visibility maintenance problem} (VMP) using the notion of controlled
invariance.  The key idea is to interpret the nonlinear model describing
the relative dynamics of the leader and the follower, as a linear system
with model parameter uncertainty, with the control input of the leader
playing the role of an external disturbance.  The~VMP can then be easily
reformulated as \emph{linear constrained regulation problem with additive
  disturbances} \mbox{(DLCRP)}. 
Positive \mbox{$\mathcal{D}$-invariance} conditions for general linear uncertain systems with
parametric disturbance matrix are introduced and used to solve the VMP when
box bounds on the state, control inputs and disturbances are considered.
Analytical conditions for the solution of the VMP are obtained by
symbolically solving with the \emph{Fourier-Motzkin elimination} method,
the set of linear inequalities defining the polytope of all the feasible
state feedback matrices.  The proposed design procedure can be easily
adapted to provide the control with \emph{unknown but bounded} (UBB)
disturbances rejection capabilities. Other extensions are also discussed:
we present conditions for the solution of the VMP when robots' desired
displacement is defined through angular parameters instead of distances,
and extend the results valid for a leader-follower pair of robots to chains
of \mbox{$n$ vehicles}.
Extensive simulation results illustrate the theory in the different
working \mbox{scenarios}. 
The present paper builds upon~\cite{MoBuPr_CDC08}, compared to which we
provide herein a more detailed and extended theory, as well as a more
accurate numerical validation.

\vspace{-0.25cm}
\subsection{Organization}
\vspace{-0.25cm}

The rest of the paper is organized as follows.
In Sect.~\ref{SEC:Prel} the linear constrained regulation problem
is reviewed and new positive $\mathcal{D}$-invariance conditions
for linear systems with parameter uncertainty are presented.
In~Sect.~\ref{SEC:VMP} we introduce the VMP and prove the main
result of the paper. In Sect.~\ref{SEC:Extensions} we investigate some extensions 
of the basic setup of~Sect.~\ref{SEC:VMP}. 
In Sect.~\ref{SEC:simul}, simulation results 
are presented. In~Sect.~\ref{SEC:ConcFut} the main contributions of the paper are summarized 
and possible avenues for future research are highlighted. 

\vspace{-0.2cm}
\section{The linear constrained regulation problem}\label{SEC:Prel} 
\vspace{-0.2cm}

This section presents a series of results that are \mbox{instrumental} in
addressing the visibility maintenance problem in Sect.~\ref{SEC:VMP}.
Our exposition will basically follow~\cite{Blanchini_JOTA91}:
Theorem~\ref{Theo3}, Corollary~\ref{Cor2}, Theorem~\ref{Theo4} and
Corollary~\ref{Cor3} extend the corresponding results in
\cite{Blanchini_TAC90,Blanchini_JOTA91}
(see~also~\cite[Ch.~4]{book_BlanchiniMi08}), to linear uncertain systems
with parametric disturbance matrix. 
Consider the following system,
\begin{equation}\label{EQ:sys_inc}
\dot{s}(t) = A(q(t))\,s(t) + B(q(t))\,u(t),
\end{equation}
where $s(t) \in \mathcal{X} \subset \mathbb{R}^n$ and $u(t) \in
\mathcal{U} \subset \mathbb{R}^m$ are respectively the state and
input vectors, $q(t) \in \mathcal{Q} \subset \mathbb{R}^p$ is the
model parameter uncertainty vector, while $\mathcal{U}$,
$\mathcal{X}$, $\mathcal{Q}$ are assigned sets containing the
origin, with $\mathcal{U}$ and $\mathcal{Q}$ compact. We assume
that $A(q)$ and $B(q)$ are matrices of suitable dimensions whose
entries are continuous functions of $q$. We will suppose $q(t)$ to
be a piecewise continuous function of time.

\begin{definition}[Positive invariance]\label{Def:PosInv}
The set $\mathcal{S} \subset \mathbb{R}^n$ is positively invariant
for system~(\ref{EQ:sys_inc}), if and only~if, for every initial
condition $s(0) \in \mathcal{S}$ and every admissible $q(t) \in
\mathcal{Q}$, the solution obtained for $u(t) \equiv 0$, satisfies
the condition $s(t) \in \mathcal{S}$ for $t > 0$.
\end{definition}

\begin{definition}[Admissible region]
A region $\mathcal{S} \subset \mathbb{R}^n$ is said to be
\emph{admissible} for the feedback control law $u=K s$, if and
only if, for every $s \in \mathcal{S}$, the condition $u \in
\mathcal{U}$ holds. If\, $\mathcal{U}$ and $\mathcal{S}$ are convex
polyhedral sets containing the origin, the admissibility of
$\mathcal{S}$ is simply equivalent~to
\begin{equation}\label{EQ:admiss}
K\text{v}_i \in \mathcal{U},\; \text{v}_i \in
\text{vert}(\mathcal{S}),\;\, i \in \{1,\ldots,\mu\},
\end{equation}
where $\text{vert}(\mathcal{S})$ denotes the set of vertices of
$\mathcal{S}$.
\end{definition}

We can now introduce the \emph{linear constrained
regulation problem} (LCRP), \cite{Blanchini_JOTA91}. 
\begin{problem}[LCRP]
\label{PROB_LCRP} Given a system in the form (\ref{EQ:sys_inc}),
find a linear feedback control law $u(t) = K s(t)$ 
and a set $\mathcal{S} \subset \mathcal{X}$ such that, for every
initial condition $s(0) \in \mathcal{S}$ and every admissible
function $q(t) \in \mathcal{Q}$, the conditions $s(t) \in
\mathcal{X}$ and $u(t) \in \mathcal{U}$ are fulfilled for $t > 0$.
\end{problem}
\begin{theorem}\label{TheoLCRP}
The LCRP has a solution if and only if there exists a feedback
matrix $K$ and a set $\mathcal{S} \subset \mathcal{X}$ that is
positively invariant and admissible for the closed loop system
\begin{equation}\label{EQ:sys_cl}
\dot{s}(t) = F(q(t))\,s(t),\vspace{0.1cm}\\
\end{equation}
where $F(q(t)) = A(q(t)) + B(q(t))K$.
\end{theorem}


\begin{theorem}[Sub-tangentiality condition]\label{Theo_subtan}
Let \mbox{$\mathcal{S} \subset \mathbb{R}^n$} be a compact and
convex set with nonempty interior. The positive invariance of
$\mathcal{S}$ for (\ref{EQ:sys_inc}) is equivalent to the
following condition: for every $s_0 \in \partial \mathcal{S}$ and
$q \in \mathcal{Q}$, 
\begin{equation}\label{EQ:con_nonlin1}
A(q)\, s_0 \in T_{\mathcal{S}}(s_0),
\end{equation}
where $T_{\mathcal{S}}(s_0)$ is the tangent cone to $\mathcal{S}$
at $s_0$ (see~\cite[Def.~3.1]{Blanchini_AUTO99}, and 
\cite[Ch.~4]{book_AubinFr90}, \cite[Ch.~5]{book_AubinCe84}) for more details).
\end{theorem}

The main difficulty in exploiting condition (\ref{EQ:con_nonlin1})
to study the positive invariance of an assigned region
$\mathcal{S}$ is that it has to be checked on the boundary of
$\mathcal{S}$. However, if convex polyhedral sets are considered,
only their vertices must be taken into account and easy algebraic
conditions can be derived. In this respect, let us consider a
system of the form~(\ref{EQ:sys_inc}), with
\vspace{-0.1cm} 
\begin{equation}\label{Eq:AqBq}
A(q(t)) = A_0 + \sum_{l=1}^p A_l\, q_l(t),\; B(q(t)) = B_0 + \sum_{l=1}^p
B_l\, q_l(t),
\end{equation}
where $A_l$ and $B_l$, $l \in \{1,\ldots,p\}$, are
constant matrices of appropriate dimension and $q(t)$ takes values
in a compact and convex polyhedron $\mathcal{Q} \subset
\mathbb{R}^p$, ($q_l(t)$ denotes the $l$-th component of vector $q(t)$). 
Let~the set $\mathcal{U}$ be compact, convex and
polyhedral as well. We consider a candidate compact and convex
polyhedral set $\mathcal{S}$ containing the origin in its interior
and we search for a feedback matrix $K$ that assures the positive
invariance of $\mathcal{S}$ for the closed loop system
(\ref{EQ:sys_cl}), (note that the 
previous assumptions on $\mathcal{S}$ will be retained throughout this 
section). Since $\mathcal{S}$ is polyhedral, then
condition (\ref{EQ:con_nonlin1}) is fulfilled on $\partial
\mathcal{S}$ if and only if is fulfilled on every \mbox{vertex of $\mathcal{S}$}.

\begin{theorem}\label{Theo_blanc}
The set $\mathcal{S}$ is positively invariant for system
(\ref{EQ:sys_cl}) with feedback $u = K s$, if and only if, for all
$\text{v}_i \in \text{vert}(\mathcal{S})$ and $w_j \in
\text{vert}(\mathcal{Q})$\,:
\begin{center}
\;\;$F(w_j)\, \text{v}_i \in T_{\mathcal{S}}(\text{v}_i),\;\; i
\in \{1,\ldots,\mu\}$, $j \in \{1,\ldots,\nu\}$.
\end{center}
\end{theorem}
\vspace{0.2cm}

The LCRP as formulated in Problem~\ref{PROB_LCRP} does not require
the stability. However, a desirable property is the global uniform stability 
of the closed loop system. 
The link between the stability property and the existence
of positively invariant regions is established by Theorem~5.2 in~\cite{Blanchini_JOTA91}.\\
%
%
Let us now turn our attention to systems in the form
\begin{equation}\label{EQ:sys_gener}
\dot{s}(t) = A(q(t))\,s(t) + B(q(t))\,u(t) + E(q(t))\,\delta(t),
\end{equation}
where the unknown external disturbance $\delta(t)$ is constrained
in a compact and convex polyhedral set $\mathcal{D} \subset
\mathbb{R}^l$ containing the origin. Note that with respect to the
systems considered in~\cite{Blanchini_JOTA91}, the structure in
(\ref{EQ:sys_gener}) is more general inasmuch as matrix $E$ also depends
on the uncertain parameter~$q$. As an immediate extension of the
positive invariance property introduced in
Definition~\ref{Def:PosInv}, we may require that the state~$s$
remains in $\mathcal{S}$ despite the presence of the
disturbance~$\delta(t)$.
\begin{definition}[Positive $\mathcal{D}$-invariance]
The set $\mathcal{S} \subset \mathbb{R}^n$ is positively
$\mathcal{D}$-invariant for system (\ref{EQ:sys_gener}), if
for every initial condition $s(0) \in \mathcal{S}$ and all
admissible $q(t) \in \mathcal{Q}$ and \mbox{$\delta(t) \in
\mathcal{D}$}, the solution obtained for $u(t) \equiv 0$,
satisfies the condition $s(t) \in \mathcal{S}$ for $t > 0$.
\end{definition}

We can now introduce the \emph{linear constrained regulation
problem with additive disturbances} (DLCRP).
\begin{problem}[DLCRP]
\label{PROB_DLCRP}
Given a system in the form (\ref{EQ:sys_gener}), find a linear feedback
control law $u(t) = K s(t)$ 
and a set $\mathcal{S} \subset \mathcal{X}$ such that, for every
initial condition $s(0) \in \mathcal{S}$ and every admissible
$q(t) \in \mathcal{Q}$ and $\delta(t) \in \mathcal{D}$, the
conditions $s(t) \in \mathcal{X}$ and $u(t) \in \mathcal{U}$ are
fulfilled for $t > 0$.
\end{problem}
\begin{theorem}\label{TheoDLCRP}
The DLCRP has a solution if and only if there exists a feedback
matrix $K$ and a set $\mathcal{S} \subset \mathcal{X}$ that is positively
$\mathcal{D}$-invariant and admissible for the closed loop system $\dot{s}(t)
= F(q(t))\,s(t) + E(q(t))\,\delta(t)$.
\end{theorem}
Similarly to~(\ref{Eq:AqBq}), we will henceforth suppose
that $E(q(t)) = E_0 + \sum_{l=1}^p E_l\, q_l(t)$.
We are now ready to state the \emph{main theorem} of this section.
\begin{theorem}[Main result]\label{Theo3} The set $\mathcal{S}$ is
positive $\mathcal{D}$-invariant for system~(\ref{EQ:sys_gener})
with feedback $u = K s$, \mbox{if and only if}, for all $\text{v}_i \in
\text{vert}(\mathcal{S})$, $\omega_j \in \text{vert}(\mathcal{Q})$
and $r_k~\in~\text{vert}(\mathcal{D})$,
\begin{equation}\label{EQ:main_theoF}
\begin{array}{l}
\!\!\!F(w_j)\,\text{v}_i + E(w_j)\,r_k \in
T_{\mathcal{S}}(\text{v}_i),\vspace{0.1cm}\\ 
\quad i \in \{1,\ldots,\mu\},\; j \in \{1,\dots,\nu\},\;k \in \{1,\dots,\eta\}.
\end{array}
\end{equation}
\end{theorem}
\vspace{-0.5cm}
\emph{Proof:} 
The proof is based on the same ideas as those in 
\cite[Th.~2.1]{Blanchini_TAC90} and 
\cite[Th.~4.1]{Blanchini_JOTA91} (see
also~\cite[Ch. 2, Sect.~4]{book_AubinCe84}) 
and the references therein).
For the \emph{necessity}, we have to prove that if
$\mathcal{S}$ is a positive $\mathcal{D}$-invariant region for
system (\ref{EQ:sys_gener}), then condition (\ref{EQ:main_theoF})
holds. The proof is straightforward, since for the
sub-tangentiality condition the positive $\mathcal{D}$-invariance
of $\mathcal{S}$ for system (\ref{EQ:sys_gener}) is equivalent to
\begin{equation}\label{EQ:subtang}
F(q)\,s + E(q)\,\delta \in T_{\mathcal{S}}(s),\; s \in
\partial\mathcal{S},\, q \in \mathcal{Q},\,\delta \in \mathcal{D},
\end{equation}
that trivially implies condition (\ref{EQ:main_theoF}).
For \emph{sufficiency}, let us consider $s$ arbitrary in
$\mathcal{S}$, $q$ arbitrary in $\mathcal{Q}$ and $\delta$
arbitrary in~$\mathcal{D}$. Supposing condition
(\ref{EQ:main_theoF}) holds, inclusion (\ref{EQ:subtang}) has to be
proved. We~have that
$s = \sum_{i=1}^{\mu} \alpha_i$,  $\text{v}_i,\;\; q =
\sum_{j=1}^{\nu} \beta_j\, w_j$, $\delta = \sum_{k=1}^{\eta}
\rho_k\, r_k$ with
$\sum_{i=1}^{\mu} \alpha_i = 1$, $\sum_{j=1}^{\nu} \beta_j = 1$, $\sum_{k=1}^{\eta} \rho_k = 1$,
for some $0 \leq \alpha_i \leq
1$,\, $i \in \{1,\ldots,\mu\}$, $0 \leq \beta_j \leq 1$, $j \in \{1,\ldots,\nu\}$ 
and $0 \leq \rho_k \leq 1$, $k \in \{1,\ldots,\eta\}$. We first prove that
\begin{equation}\label{EQ:partial_step}
\begin{array}{c}
\hspace{-2.5cm} F(q)\,\text{v}_i + E(q)\,r_k \in T_{\mathcal{S}}(\text{v}_i),\vspace{0.15cm}\\
\qquad\quad  i \in \{1,\ldots,\mu\},\; q \in \mathcal{Q},\;k \in
\{1,\dots,\eta\}.
\end{array}
\end{equation}
Let $w_{lj}$ be the $l$-th entry of $w_j$. We have that
\begin{align*}
& F(q)\,\text{v}_i + E(q)\,r_k = \vspace{0.02cm}\\ 
& \!\!=\! \big(F_0\! +
\sum_{l=1}^p F_l\, q_l\big) \text{v}_i + \big(E_0 + \sum_{l=1}^p
E_l\, q_l\big) r_k \vspace{0.02cm}\\ & \!\!=\! \big(F_0 +\!
\sum_{l=1}^p F_l \sum_{j=1}^{\nu} \beta_j\, w_{lj}\big) \text{v}_i
+ \big(E_0 + \!\sum_{l=1}^p E_l \sum_{j=1}^{\nu} \beta_j\,
w_{lj}\big) r_k \vspace{0.04cm}\\ & \!\!=\! \sum_{j=1}^{\nu}
\beta_j \big[\big(F_0 +\! \sum_{l=1}^p\! F_l\, w_{lj}\big)
\text{v}_i\big] \!+\!\! \sum_{j=1}^{\nu}
\beta_j\big[\big(E_0 \!+\!\! \sum_{l=1}^p\! E_l\, w_{lj}\big) r_k\big] \vspace{0.04cm}\\
& \!\!= \!\sum_{j=1}^{\nu} \beta_j \big[F(w_j)\, \text{v}_i
+ E(w_j)\, r_k \big],\vspace{0.02cm}\\ & \hspace{3cm} i \in \{1,\ldots,\mu\},\; k \in
\{1,\ldots,\eta\}.
\end{align*}
From (\ref{EQ:main_theoF}) we have that
$\sum_{j=1}^{\nu} \beta_j \big[F(w_j)\, \text{v}_i + E(w_j)\, r_k
\big] \in\, T_{\mathcal{S}}(\text{v}_i)$, 
$i \in \{1,\ldots,\mu\}$, $k \in \{1,\dots,\eta\}$, 
therefore (\ref{EQ:partial_step}) is proved. If $\pi_i$ is a
delimiting plane of $\mathcal{S}$ for~$s$ (i.e., such that $g_i^T s
= \xi_i$), we may write $s$ as a convex combination of the
vertices of $\mathcal{S}$ that belong to $\pi_i$: 
$s = \sum_{h=1}^{\mu_i} \alpha_h\,\text{v}_h$, 
with $g_i^T \text{v}_h = \xi_i$ and
$\sum_{h=1}^{\mu_i} \,\alpha_h = 1$,  $0 \leq \alpha_h
\leq 1$, $h \in \{1,\ldots,\mu_i\}$. 
Then $g_i^T(F(q) s + E(q)\,\delta) = g_i^T(F(q) \!\sum_{h=1}^{\mu_i}
\alpha_h\,\text{v}_h + E(q) \!\sum_{k=1}^{\eta} \rho_k\,r_k)$.
But from~(\ref{EQ:partial_step}) and recalling the expression of
the tangent cone when $\mathcal{S}$ is described in terms of its
vertices, we have that $g_i^T(F(q)\,\text{v}_h + E(q)\,r_k) \leq 0$, 
that implies $g_i^T\big(F(q) \sum_{h=1}^{\mu_i} \alpha_h
\text{v}_h + E(q) \sum_{k=1}^{\eta} \rho_k\,r_k\big) \leq 0$.
By~considering all the planes for $s$, condition
(\ref{EQ:subtang}) follows \mbox{immediately}. \hfill$\blacksquare$
%
The application of Theorem~\ref{Theo3} requires the knowledge of
all cones $T_{\mathcal{S}}(\text{v}_i)$, $i \in \{1,\ldots,\mu\}$.
An alternative solution is given by the following corollary (whose proof is analogous to that
of~\cite[Corollary 4.1]{Blanchini_JOTA91}) in
which the Euler auxiliary system associated to~(\ref{EQ:sys_gener}) is
involved (cf. \cite[Sect.~12.1]{book_BlanchiniMi08}).
\begin{corollary}\label{Cor2}
The set $\mathcal{S}$ is positively $\mathcal{D}$-invariant for
system~(\ref{EQ:sys_gener}), if and only if there exists $\tau >
0$ such that, for all ${\text{v}_i \in\text{vert}(\mathcal{S})}$,
$\omega_j \in \text{vert}(\mathcal{Q})$ and $r_k \in
\text{vert}(\mathcal{D})$,
\begin{equation}\label{EQ:Corol}
\begin{array}{l}
\!\!\!\!\text{v}_i + \tau(F(w_j)\,\text{v}_i + E(w_j)\,r_k) \in
\mathcal{S},\vspace{0.08cm}\\  
\quad i \in \{1,\ldots,\mu\},\; j \in \{1,\dots,\nu\},\;k \in \{1,\dots,\eta\}.
\end{array}
\end{equation}
\end{corollary}
%
%
\vspace{-0.35cm}
To~overcome the problem of the choice of $\tau$, we introduce
Theorem~\ref{Theo4} that provides a condition equivalent to~(\ref{EQ:Corol}). 
The~proof is analogous to that of \cite[Th.~2.3]{Blanchini_TAC90}.
Let $\mathcal{C}_i$ be the convex cone defined by the 
delimiting planes of $\mathcal{S}$ which contain~$\text{v}_i$
(see~\cite[Ch.~4]{Book_Panik93})\,:\vspace{-0.15cm}
$$
\begin{array}{l}
\hspace{-0.5cm}\mathcal{C}_i = \{g_h^T s \leq \xi_h,\,
\xi_h > 0,\, \text{for}\; \text{every}\;g_h^T\vspace{0.08cm}\\
\hspace{1.5cm}\text{and}\;\xi_h : {g_h^T \text{v}_i = \xi_h},\, \text{v}_i
\in \text{vert}(\mathcal{S})\}.
\end{array}
$$
\begin{theorem}\label{Theo4}
The set $\mathcal{S}$ is positively $\mathcal{D}$-invariant for
system~(\ref{EQ:sys_gener}), if and only if, for all $\tau >0$,
$\text{v}_i \in \text{vert}(\mathcal{S})$, $\omega_j \in
\text{vert}(\mathcal{Q})$ and $r_k \in
\text{vert}(\mathcal{D})$: 
$$
\begin{array}{l}
\hspace{-0.1cm}\text{v}_i + \tau(F(w_j)\,\text{v}_i
+ E(w_j)\,r_k) \in \mathcal{C}_i,\vspace{0.08cm}\\
\hspace{1cm} i \in \{1,\ldots,\mu\},\;\, j
\in \{1,\dots,\nu\},\;\, k \in \{1,\dots,\eta\}.
\end{array}
$$
\end{theorem}
\vspace{-0.45cm}
If the plane description of $\mathcal{S}$ is available, the next
corollary, whose proof directly follows from that of
\mbox{Theorem~\ref{Theo3}}, holds.
\begin{corollary}\label{Cor3}
The set $\mathcal{S}$ is positively $\mathcal{D}$-invariant for
system (\ref{EQ:sys_gener}), if and only if, for every $\tau >0$
and every $\text{v}_i \in \text{vert}(\mathcal{S})$, $\omega_j \in
\text{vert}(\mathcal{Q})$, 
\begin{equation}\label{EQ:Cor_final}
(I_n + \tau\,F(w_j))\,\text{v}_i \in
\mathcal{C}^{\star}_i,\;\, i \in \{1,\ldots,\mu\},\, j \in \{1,\dots,\nu\},
\end{equation}
where $\mathcal{C}_i^{\star}$ is the cone obtained by shifting the
planes of $\mathcal{C}_i$ as follows\,:
\vspace{-0.1cm}
$$
\begin{array}{l}
\!\mathcal{C}_i^{\star} = \{g_h^T s \leq \xi_h \,-\, \max_{jk}\{\tau
g_h^T E(w_j)\, r_k\},\,\omega_j \in \text{vert}(\mathcal{Q}),\vspace{0.1cm}\\ \hspace{1.5cm} r_k
\in \text{vert}(\mathcal{D}),\, \text{for}\;
\text{every}\;g_h^T\!:g_h^T \text{v}_i = \xi_h\}.
\end{array}
$$
\end{corollary}

\begin{remark}
According to Theorem~\ref{TheoDLCRP}, conditions
(\ref{EQ:Cor_final}) and~(\ref{EQ:admiss}) provide us with a set
of inequalities in the unknown $K$ defining the polytope
$\mathcal{K}$ of all the state feedback matrices solving the
DLCRP.
\end{remark}

\section{The visibility maintenance problem}
\label{SEC:VMP}

\subsection{Modeling}\label{SUBSEC:Mod}

\begin{figure}[t!]
  \psfrag{a}{$\subscr{x}{0}$}
  \psfrag{b}{$\subscr{y}{0}$}
  \psfrag{c}{$\subscr{\sum}{0}$}
  \psfrag{d}{$\subscr{x}{F}$}
  \psfrag{e}{$\subscr{y}{F}$}
  \psfrag{f}{$\subscr{\theta}{F}$}
  \psfrag{g}{$\subscr{\sum}{F}$}
  \psfrag{i}{$\subscr{x}{L}$}
  \psfrag{h}{$\subscr{y}{L}$}
  \psfrag{l}{$\subscr{\theta}{L}$}
  \psfrag{m}{$\subscr{\sum}{L}$}
\begin{center}
           \includegraphics[width=.83\columnwidth]{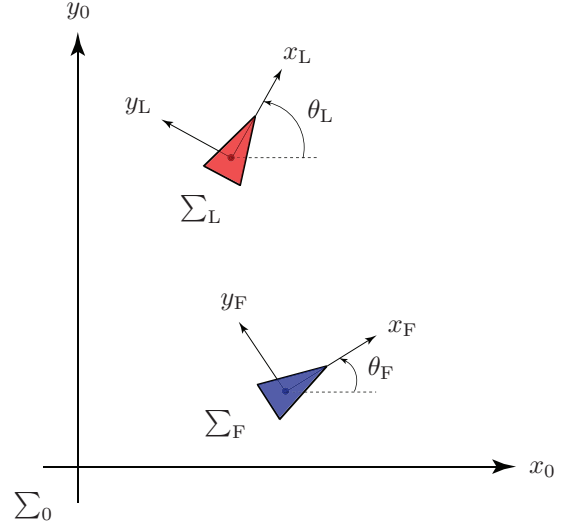}
           \caption{Leader-follower setup.}
           \label{FIG:Setup_robot}
\end{center}
\end{figure}

Let $\subscr{\Sigma}{0} \equiv \{\subscr{O}{0}\,; \subscr{x}{0},\,
\subscr{y}{0}\}$ be the fixed reference frame in~$\mathbb{R}^2$,
and $\subscr{\Sigma}{F} \equiv \{\subscr{O}{F}\,; \subscr{x}{F},\,
\subscr{y}{F}\}$ and $\subscr{\Sigma}{L} \equiv \{\subscr{O}{L}\,;
\subscr{x}{L},\, \subscr{y}{L}\}$ the reference frames attached to
a \emph{follower} robot $\textup{F}$ and a \emph{leader} robot
$\textup{L}$ (see~Fig.~\ref{FIG:Setup_robot}). The robots are
supposed to have single integrator dynamics,
\begin{equation}\label{EQ:int_dyn}
\begin{array}{cc}
\begin{array}{c}
\subupscr{\dot{p}}{\,F}{\,F} = \subupscr{\sigma}{F}{F}\,,\vspace{0.15cm}\\
\subscr{\dot{\theta}}{\,F} = \subscr{\omega}{F}\,,
\end{array}
\;&\;
\begin{array}{c}
\subupscr{\dot{p}}{\,L}{\,L} = \subupscr{\sigma}{L}{L}\,,\vspace{0.15cm}\\
\subscr{\dot{\theta}}{\,L} = \subscr{\omega}{L}\,,
\end{array}
\end{array}
\end{equation}
where $\subupscr{p}{\,F}{\,F} = (\subscr{x}{F},\,
\subscr{y}{F})^T$, $\subupscr{p}{\,L}{\,L} = (\subscr{x}{L},\,
\subscr{y}{L})^T$ are the positions, $\subupscr{\sigma}{F}{F} =
(\subupscr{\sigma}{F}{F}[1], \subupscr{\sigma}{F}{F}[2])^T$,
$\subupscr{\sigma}{L}{L} = (\subupscr{\sigma}{L}{L}[1],
\subupscr{\sigma}{L}{L}[2])^T$ the linear velocities and
$\subscr{\omega}{F}$, $\subscr{\omega}{L}$ the angular velocities
of robots $\textup{F}$ and~$\textup{L}$ in the frames
$\subscr{\Sigma}{F}$ and $\subscr{\Sigma}{L}$, respectively. We
are going to derive a dynamic model describing the relative
dynamics of the robots $\textup{F}$ and~$\textup{L}$. Referring
(\ref{EQ:int_dyn}) to the frame $\subscr{\Sigma}{0}$, we obtain~\cite{Book_SicilianoScViOr08},
$\subupscr{\dot{p}}{F}{0} = \subupscr{R}{F}{0}(\subscr{\theta}{F})\,\subupscr{\sigma}{F}{F}$,
$\subupscr{\dot{p}}{L}{0}
= \subupscr{R}{L}{0}(\subscr{\theta}{L})\,
\subupscr{\sigma}{L}{L}$,
where 
$$
\!\subupscr{R}{F}{0}(\subscr{\theta}{F}) \!=\!
\left[\begin{matrix}\cos\subscr{\theta}{F} &
-\sin\subscr{\theta}{F}\\\sin\subscr{\theta}{F} &
\cos\subscr{\theta}{F}
\end{matrix}\right]\!,\, \subupscr{R}{L}{0}(\subscr{\theta}{L}) \!=\!
\left[\begin{matrix}\cos\subscr{\theta}{L} &
-\sin\subscr{\theta}{L}\\\sin\subscr{\theta}{L} &
\cos\subscr{\theta}{L}
\end{matrix}\right]\!\!.
$$ 
\noindent The position of robot $\textup{L}$ with respect to
$\subscr{\Sigma}{F}$ is then given~by
\begin{equation}\label{Eq:p^F_L}
\subupscr{p}{L}{F} = \subupscr{R}{0}{F}(\subscr{\theta}{F})(\subupscr{p}{L}{0} -
\subupscr{p}{F}{0}),\vspace{0.16cm}
\end{equation}
where $\subupscr{R}{0}{F}(\subscr{\theta}{F}) = (\subupscr{R}{F}{0}(\subscr{\theta}{F}))^T$.
Differentiating (\ref{Eq:p^F_L}) with respect to time, we get
\begin{equation}\label{EQ:rel_pos_diff}
\subupscr{\dot{p}}{\,L}{\,F} \!=
\subupscr{\dot{R}}{0}{F}(\subscr{\theta}{F})(\subupscr{p}{L}{0} -
\subupscr{p}{F}{0}) +
\subupscr{R}{0}{F}(\subscr{\theta}{F})\big(\subupscr{R}{L}{0}(\subscr{\theta}{L})\,
\subupscr{\sigma}{L}{L} -
\subupscr{R}{F}{0}(\subscr{\theta}{F})\,\subupscr{\sigma}{F}{F}\,\big).
\end{equation}
Since
$$
\subupscr{\dot{R}}{0}{F}(\subscr{\theta}{F}) =
\left[\!\begin{smallmatrix} 0 \,& \subscr{\omega}{F}
\vspace{0.1cm}\\
-\subscr{\omega}{F} \,&
0\end{smallmatrix}\!\right]\!\subupscr{R}{0}{F}(\subscr{\theta}{F}),
$$
we can rewrite (\ref{EQ:rel_pos_diff})~as
\begin{equation}\label{EQ:rel_pos_diff2}
\subupscr{\dot{p}}{L}{F} = \left[\!\begin{array}{cc} 0 &
\subscr{\omega}{F}\\-\subscr{\omega}{F} &
0\end{array}\right] \subupscr{p}{L}{F} +
\subupscr{R}{L}{F}(\subupscr{\beta}{L}{F})\, \subupscr{\sigma}{L}{L} -
\subupscr{\sigma}{F}{F},
\end{equation}
where $\subupscr{\beta}{L}{F} \triangleq \subscr{\theta}{L} - \subscr{\theta}{F}$. 
Collecting equation~(\ref{EQ:rel_pos_diff2}) and the relative angular
dynamics of the robots together, \mbox{we obtain} the following system
\begin{equation}\label{EQ:nonlin}
\!\left[
\begin{MAT}[4pt]{c}
\subupscr{\dot{p}}{L}{F}\vspace{0.3cm}\\
\dot{\subupscr{\beta}{L}{F}} \addpath{(0,1,0)l}\\
\end{MAT}\right] \!\!=\!\!
\left[
\begin{array}{c|cc}
-I_2 \;&\; \begin{matrix} \subupscr{p}{L}{F}[2]\vspace{0.15cm}\\
-\subupscr{p}{L}{F}[1]\vspace{0.1cm}\end{matrix}\\
\hline \begin{matrix} 0\; &\;
0\end{matrix} \;&\; - 1
\end{array}
\right]\!\!\! \left[
\begin{MAT}[4pt]{c}
\subupscr{\sigma}{F}{F}\vspace{0.4cm}\\
\subscr{\omega}{F} \addpath{(0,1,0)l}\\
\end{MAT}\right]
+ \left[
\begin{array}{c|cc}
\subupscr{R}{L}{F}(\subupscr{\beta}{L}{F}) \,&\; \begin{matrix} 0\vspace{0.15cm}\\
0\vspace{0.1cm}\end{matrix}\\
\hline\begin{matrix} 0\; &\; 0\end{matrix} \,&\; 1
\end{array}
\right]\!\!\!\left[
\begin{MAT}[4pt]{c}
\subupscr{\sigma}{L}{L}\vspace{0.4cm}\\
\subscr{\omega}{L} \addpath{(0,1,0)l}\\
\end{MAT}\right]\!\!,
\end{equation}
where $\subupscr{p}{L}{F}=(\subupscr{p}{L}{F}[1]$,
$\subupscr{p}{L}{F}[2])^T$. For the sake of simplicity, we will
suppose that
\begin{equation}\label{EQ:dubin}
\begin{array}{c}
\subupscr{\sigma}{F}{F} = (1 + \subscr{v}{F}\,,\; 0)^T\!,\vspace{0.2cm}\\
\subupscr{\sigma}{L}{L} = (1 + \subscr{v}{L}\,,\; 0)^T\!,
\end{array}
\end{equation}
where $|\subscr{v}{F}(t)| < 1$, $|\subscr{v}{L}(t)| < 1$, for all
$t \geq 0$. $\textup{F}$ and $\textup{L}$ will then behave in a way similar to \emph{Dubins 
vehicles} since they can only move forward (however, differently 
from the standard Dubins model, $\subscr{v}{F}$ and $\subscr{v}{L}$ are not necessarily 
constant in our~case).
Substituting (\ref{EQ:dubin}) in (\ref{EQ:nonlin}), we
finally come up with the following system
\begin{equation}\label{EQ:nonlin_dubin}
\begin{array}{c}
\left[
\begin{array}{c}
\subupscr{\dot{p}}{L}{F}[1]\vspace{0.17cm}\\
\subupscr{\dot{p}}{L}{F}[2]\vspace{0.17cm}\\
\dot{\subupscr{\beta}{L}{F}}
\end{array}\right]  =  \left[\,
\begin{matrix}
\cos\subupscr{\beta}{L}{F} - 1\vspace{0.15cm}\\
\sin\subupscr{\beta}{L}{F}\vspace{0.15cm}\\
0
\end{matrix}\right] + \left[
\begin{matrix}
-1& \subupscr{p}{L}{F}[2]\vspace{0.15cm}\\
0 & -\subupscr{p}{L}{F}[1]\vspace{0.15cm}\\
0 & -1
\end{matrix}\right]\!
\left[\,
\begin{matrix}
\subscr{v}{F}\vspace{0.08cm}\\
\subscr{\omega}{F}
\end{matrix}\,\right]\vspace{0.25cm}\\ 
+ \left[\,
\begin{matrix}
\cos\subupscr{\beta}{L}{F} & 0\vspace{0.15cm}\\
\sin\subupscr{\beta}{L}{F} & 0\vspace{0.15cm}\\
0 & 1
\end{matrix}\,\right]\!
\left[\,
\begin{matrix}
\subscr{v}{L}\vspace{0.08cm}\\
\subscr{\omega}{L}
\end{matrix}\,\right]\!,
\end{array}
\end{equation}
%
%
with state vector $s = (\subupscr{p}{L}{F}[1],\subupscr{p}{L}{F}[2],\subupscr{\beta}{L}{F})^T \in
\mathcal{X} \subset \textup{SE}(2)$, input 
vector $u = (\subscr{v}{F},\, \subscr{\omega}{F})^T \in
\mathcal{U} \subset (-1,\, 1) \times \mathbb{R}$ and disturbance
vector $\delta = (\subscr{v}{L},\,\subscr{\omega}{L})^T \in
\mathcal{D} \subset (-1,\, 1) \times~\mathbb{R}$. \mbox{System}~(\ref{EQ:nonlin_dubin}) 
describes the relative dynamics of the Dubins-like vehicles $\textup{F}$ and $\textup{L}$ 
in the configuration space~$\textup{SE}(2)$. 

\subsection{Problem statement}\label{SUBSEC:Prob}

In the forthcoming analysis, we will suppose that robot
$\textup{F}$ is equipped with a sensor (e.g., a panoramic camera,
a laser range finder, etc.) with limited sensing range. We~will
call \emph{visibility set} of robot $\textup{F}$ any compact and
convex polyhedral set $\mathcal{S} \subset \mathcal{X}$ containing
the origin in its interior.
Please note that the visibility set generalizes the notion of
\emph{sensor footprint} since it encodes both the position and
angle \mbox{information}.\\
Robot $\textup{L}$ moves along a given
\mbox{trajectory} and robot $\textup{F}$ aims at keeping
$\textup{L}$ always inside its visibility set $\mathcal{S}$, while
respecting the control bounds. 
By~referring to system (\ref{EQ:nonlin_dubin}), we can formalize
this problem as follows: 
\begin{problem}[\footnotesize{Visibility maintenance problem:} \scriptsize{VMP}]
  \label{PROB_vis_main} Let~$\mathcal{S}$ be the visibility set of robot
  $\textup{F}$ and let $s(0) \in \mathcal{S}$. Find a control $u(t)$ such
  that for all $\delta(t) \in \mathcal{D}$, the conditions $s(t) \in
  \mathcal{S}$ and $u(t) \in \mathcal{U}$ are fulfilled for $t > 0$.
\end{problem}


In the following, we will refer to Problem~\ref{PROB_vis_main} as to the VMP with
candidate positively $\mathcal{D}$-invariant set $\mathcal{S}$, 
control set $\mathcal{U}$ and disturbance set $\mathcal{D}$. 

\subsection{Solution method}\label{SUBSEC:Sol}

Next, we will transcribe system~(\ref{EQ:nonlin_dubin}) into the linear
parametric form~(\ref{EQ:sys_gener}): in this way, the VMP simply reduces
to the DLCRP (Problem~\ref{PROB_DLCRP}) introduced in Sect.~\ref{SEC:Prel} 
and suitable solvability conditions can be derived by means of
(\ref{EQ:Cor_final}) and (\ref{EQ:admiss}).\\ 
%
%
After simple matrix manipulations in~(\ref{EQ:nonlin_dubin}), we obtain
\begin{equation}\label{EQ:nonlin_unc_param}
\begin{array}{l}
\!\!\left[
\begin{matrix}\Delta\subupscr{\dot{p}}{L}{F}[1]\vspace{0.1cm}\\
\subupscr{\dot{p}}{L}{F}[2]\vspace{0.1cm}\\
\dot{\subupscr{\beta}{L}{F}}
\end{matrix} \right] = \left[\,
\begin{matrix}
0 \;&\; 0 &\; \frac{\cos\subupscr{\beta}{L}{F} - 1}{\subupscr{\beta}{L}{F}}\vspace{0.1cm}\\
0 \;&\; 0 &\; \frac{\sin\subupscr{\beta}{L}{F}}{\subupscr{\beta}{L}{F}}\vspace{0.1cm}\\
0 \;&\; 0 &\; 0
\end{matrix}\,\right]\!\!
\left[
\begin{matrix}
\Delta \subupscr{p}{L}{F}[1]\vspace{0.1cm}\\
\subupscr{p}{L}{F}[2]\vspace{0.1cm}\\
\subupscr{\beta}{L}{F}
\end{matrix} \right]\vspace{0.25cm}\\
\quad + \left[
\begin{matrix}
-1 & \subupscr{p}{L}{F}[2]\vspace{0.1cm}\\
0 & - d - \Delta \subupscr{p}{L}{F}[1]\vspace{0.1cm}\\
0 & -1
\end{matrix} \right]\!\!
\left[
\begin{matrix}
\subscr{v}{F}\vspace{0.1cm}\\
\subscr{\omega}{F}
\end{matrix} \right] + \left[\,
\begin{matrix}
\cos\subupscr{\beta}{L}{F} & 0\vspace{0.1cm}\\
\sin\subupscr{\beta}{L}{F} & 0\vspace{0.1cm}\\
0 & 1
\end{matrix}\,\right]\!\!
\left[
\begin{matrix}
\subscr{v}{L}\vspace{0.1cm}\\
\subscr{\omega}{L}
\end{matrix}\right]\!,
\end{array}
\end{equation}
which can be written in the form (\ref{EQ:sys_gener}) with
\begin{equation}\label{Eq:AqBqEq}
\begin{array}{cc}
A(q) = \left[\,\begin{matrix}
0 \;&\; 0 & q_2\vspace{-0.05cm}\\
0 \;&\; 0 & 1+q_1\vspace{-0.05cm}\\
0 \;&\; 0 & 0
\end{matrix} \right]\!,\;\;
& B(q) = \left[\begin{matrix}
-1 \;&\; q_4\vspace{-0.05cm}\\
0 \;&\; -d - q_3\vspace{-0.05cm}\\
0 \;&\; -1
\end{matrix} \right]\!,
\end{array}
\end{equation}
\vspace{-0.83cm}\\
$$
E(q) = \left[\,\begin{matrix}
1 + q_5 \;&\; 0\vspace{-0.05cm}\\
q_6 \;&\; 0\vspace{-0.05cm}\\
0 \;&\; 1
\end{matrix}\,\right]\!,
$$
\noindent where 
\begin{equation}
  \label{eq:q-definition}
\begin{array}{ccc}
\displaystyle \hspace{-0.2cm} q_1 = \frac{\sin\subupscr{\beta}{L}{F}}{\subupscr{\beta}{L}{F}} - 1,\; & 
\displaystyle q_2 = \frac{\cos\!\subupscr{\beta}{L}{F} -1}{\subupscr{\beta}{L}{F}},\; & q_3 = \Delta
\subupscr{p}{L}{F}[1],\vspace{0.2cm}\\ 
\hspace{-0.2cm} q_4 = \subupscr{p}{L}{F}[2], & q_5 =
\cos\subupscr{\beta}{L}{F} - 1, & q_6 = \sin\subupscr{\beta}{L}{F}.
\end{array}
\end{equation}
We made the following change
of variables in~(\ref{EQ:nonlin_unc_param}),
$$
(\subupscr{p}{L}{F}[1], \subupscr{p}{L}{F}[2], \subupscr{\beta}{L}{F})^T
\rightarrow\, (\Delta \subupscr{p}{L}{F}[1],
\subupscr{p}{L}{F}[2], \subupscr{\beta}{L}{F})^T,
$$
where $\Delta \subupscr{p}{L}{F}[1]
= \subupscr{p}{L}{F}[1] - d$ and $d$ is a strictly positive constant. 
Two main reasons motivated this transformation: first of all, if robot
$\textup{F}$ is able to keep~$\textup{L}$ always inside a
visibility set displaced of $d$ with respect to its center (with
$d > \max_{\,s_1,s_2 \,\in\, \text{vert}(\mathcal{S})} \,\frac{1}{2}\,\|s_1-s_2\|_2$), 
then this automatically guarantees robots'
collision avoidance. Second, this choice
simplifies the study of the VMP with chains 
of robots (see~Sect.~\ref{SEC:ext_chain}).\\
\begin{figure}[t!]
\begin{center}
\psfrag{x}{$\subscr{x}{L}$}
\psfrag{y}{$\subscr{y}{L}$}
\psfrag{m}{$\mathcal{S}$}
\includegraphics[width=0.84\columnwidth]{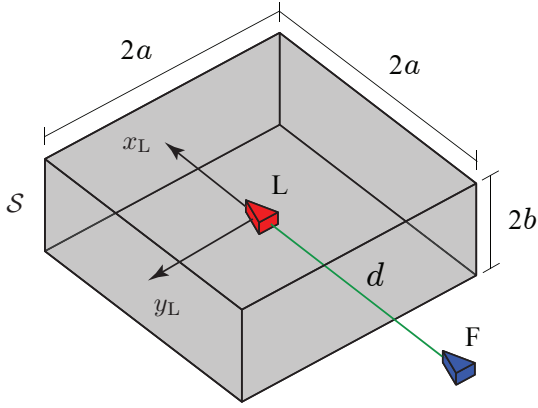}
\end{center}
\caption{The visibility set $\mathcal{S}$ in (\ref{EQ:VisBox}) and the pose of the
robots $\textup{L}$ and $\textup{F}$ for $(\Delta
\subupscr{p}{L}{F}[1], \subupscr{p}{L}{F}[2], \subupscr{\beta}{L}{F})^T =
(0,0,0)^T$, $d > a$.}\label{FIG:setS}
\end{figure}
%
Notice that $A_0$, $B_0$ and $E_0$ in (\ref{Eq:AqBqEq}) (recall the
notation used in Sect.~\ref{SEC:Prel}) correspond to the constant
matrices obtained by linearizing system (\ref{EQ:nonlin_dubin})
around the equilibrium $s_{eq} = (d,\,0,\,0)^T,$ $u_{eq} =
(0,\,0)^T$, $\delta_{eq} = (0,\,0)^T$.\\
%
%
For the sake of simplicity, we will henceforth make the following assumption: 
\begin{assumption}\label{Assumpt}
Suppose that (see Fig.~\ref{FIG:setS})
\begin{equation}\label{EQ:boundsUD}
\begin{array}{l}
\hspace{-0.1cm}\mathcal{U} = \left\{(\subscr{v}{F},\,
\subscr{\omega}{F})^T:\, \subscr{v}{F} \in [-\subscr{V}{F},\,\subscr{V}{F}],\; 
\subscr{\omega}{F} \in [-\subscr{\Omega}{F},\,\subscr{\Omega}{F}]\right\}\!,\vspace{0.2cm}\\
\hspace{-0.12cm}\mathcal{D} = \left\{(\subscr{v}{L},\,
\subscr{\omega}{L})^T:\, \subscr{v}{L} \in
[-\subscr{V}{L},\, \subscr{V}{L}],\; \subscr{\omega}{L} \in
[- \subscr{\Omega}{L},\,\subscr{\Omega}{L}]\right\}\!,
\end{array}
\end{equation}
\begin{equation}\label{EQ:VisBox}
\begin{array}{c}
\hspace{-0.82cm}\mathcal{S} = \big\{(\Delta \subupscr{p}{L}{F}[1],
\subupscr{p}{L}{F}[2], \subupscr{\beta}{L}{F})^T\!:\, \Delta
\subupscr{p}{L}{F}[1] \in [-a,\,a],\vspace{0.1cm}\\ 
\;\; \subupscr{p}{L}{F}[2] \in
[-a,\,a],\; \subupscr{\beta}{L}{F} \in [-b,\,b]\big\},
\end{array}
\end{equation}
where $\subscr{V}{F} < 1$, $\subscr{V}{L} < 1$,
$\subscr{\Omega}{F}$, $\subscr{\Omega}{L}$, $a$, $b$, are strictly
positive constants.
\end{assumption}
The definition~(\ref{EQ:boundsUD}) of control and
disturbance sets is motivated by the presence of saturation bounds on the
driving motors of physical robots. The candidate invariant set $\mathcal{S}$ is defined
as in~(\ref{EQ:VisBox}) because it is computationally simple to handle 
(this will allow us to provide concise solvability conditions for the VMP in
Theorem~\ref{Teo1}), and because its horizontal section represents a 
reasonable good inner approximation of a disk sensor footprint 
e.g., due to an omnidirectional camera or a 360$^{\circ}$ laser
scanner (the problem of precisely quantify the non-conservatism
introduced by this approximation goes beyond the scope of this 
paper, and it is left as a subject of future research). Finally,
it is worth emphasizing that is not unusual in the multi-agent systems 
literature to encounter rectangular footprints, 
that are typically used, for example, to model 
``push-broom'' or line-scanner sensors (see, e.g., \cite{FinkePaGaSp_LNCIS05}).\\
We now complete our transcription of system~(\ref{EQ:nonlin_dubin})
into the linear parametric form~(\ref{EQ:sys_gener}), by defining an
appropriate polyhedral set for the model parameter uncertainty:
\vspace{-0.2cm}
\begin{equation}\label{Eq:inter_q}
\begin{array}{l}
\!\mathcal{Q}
    = \big\{(q_1,\dots,q_6)^T:\, 
      q_1 \in \big[\frac{\sin b}{b} - 1,\,0\big],\vspace{0.245cm}\\
      \qquad\; q_2 \in \left[\frac{\cos b - 1}{b},\frac{1 - \cos
      b}{b}\right],\; q_3 \in \left[-a,\, a\right],\; 
      q_4 \in \left[-a,\, a\right],\vspace{0.23cm}\\ 
      \qquad\; q_5 \in \big[\!\cos b -1,\, 0\big],\;
      q_6 \in \left[-\sin b,\, \sin b\right]\big\}. 
\end{array}
\end{equation}
It is easy to show that, if $(\Delta\subupscr{p}{L}{F}[1],
\subupscr{p}{L}{F}[2],\subupscr{\beta}{L}{F})^T \in \,\mathcal{S}$, then
definition~\eqref{eq:q-definition} immediately implies that $q\in \mathcal{Q}$.

%
\begin{remark}
In the previous passages, the nonlinear system 
(\ref{EQ:nonlin_dubin}) has been \emph{absorbed} into a linear (controlled) 
differential inclusion (see \cite[Sect.~2.1.2]{book_BlanchiniMi08}).
This is an approximate transformation: however, no matter how the
input $u$ is chosen, we have that any trajectory of the original 
system~(\ref{EQ:nonlin_dubin}) is also a trajectory of the 
corresponding linear uncertain system (the
opposite is clearly not true in general). \mbox{As a consequence}, if we are able to
determine the qualitative behavior of the absorbing system, we can determine
(in~a conservative~way) the behavior of the original system. 
Some tools are available in the robust control literature to quantify this
conservativeness, such as, e.g., the recent nonlinear extensions of
the \emph{gap} and Vinnicombe's $\nu$-\emph{gap metrics}  (see 
\cite{BianFr_SIAMJCO05,JamesSmVi_SIAMJCO05} and the references
therein). However, in the interest of brevity, we will not perform
such an analysis in this paper.
\end{remark}
\vspace{-0.1cm}
\noindent We are now ready to state the \emph{main result} of this
section.
\vspace{-0.1cm}
\begin{theorem}[Solvability of the VMP]\label{Teo1}
  For the robots $\textup{F}$ and $\textup{L}$, consider the VMP with
  candidate positive $\mathcal{D}$-invariant set $\mathcal{S}$, 
  control set $\mathcal{U}$ and disturbance set $\mathcal{D}$ 
  satisfying Assumption~\ref{Assumpt} with $d
  > a$, $0 < b \leq \pi/2$. This VMP has a solution if the following
  conditions are satisfied \vspace{-0.15cm}
\begin{equation}\label{EQ:cond1}
\subscr{V}{F} \geq \subscr{V}{L}\left(1 + \frac{a\sin b}{d-a}\right) +\, 1 -\, \cos b\, +\, \frac{a\,b}{d-a}\,,
\end{equation}
\begin{equation}\label{EQ:cond2}
\subscr{\Omega}{L} \leq \frac{(1-\subscr{V}{L})\sin b}{d+a}\,,
\;\;\, \frac{\subscr{V}{L}\, \sin b + b}{d-a}\, \leq\,
\subscr{\Omega}{F}\,.
\end{equation}
\vspace{2mm}\\
The state feedback matrix has the form
\begin{equation}\label{Eq:feed_mat_K}
K=\left[\begin{array}{ccc}
k_{11} & 0 & 0\vspace{0.05cm}\\
0 & k_{22} & k_{23}
\end{array}
\right]\!,
\end{equation}
where $k_{11}$, $k_{22}$ and $k_{23}$ belong to the polytope
$\mathcal{K} \subset \mathbb{R}^3$ defined by
(\ref{Eq:All_cond1})-(\ref{Eq:All_cond2}), (see the proof below).
\end{theorem}
\emph{Proof:}\; Let us apply Corollary~\ref{Cor3} to
system~(\ref{EQ:nonlin_unc_param}). By~selecting $\tau = 1$ in
(\ref{EQ:Cor_final}), we obtain 
\scriptsize
\begin{equation}\label{EQ:cond1_proof}
\!\!\left[
\begin{matrix}
1-k_{11}+q_4 k_{21} & -k_{12}+ q_4 k_{22} & q_2 - k_{13} + q_4 k_{23}\vspace{0.1cm}\\
-(d+q_3)k_{21} & 1-(d+q_3) k_{22} &\, 1+q_1-(d+q_3) k_{23}\vspace{0.1cm}\\
-k_{21} & -k_{22} & 1-k_{23}
\end{matrix}
\right]\!v_i \in\, \mathcal{C}_i^{\star}.
\end{equation}
\normalsize
Condition (\ref{EQ:cond1_proof}) must be evaluated only on the
vertices $\text{v}_1 = (a,a,b)^T, \text{v}_2 = (a,a,-b)^T,
\text{v}_3 = (a,-a,b)^T, \text{v}_4 =(a,-a,-b)^T$ since the set
(\ref{EQ:VisBox}), whose plane representation~is
$$
\left[
\begin{array}{ccc}
\displaystyle 1/a & 0 & 0\vspace{0.01cm}\\
\displaystyle -1/a & 0 & 0\vspace{0.01cm}\\
0 & \displaystyle 1/a & 0\vspace{0.01cm}\\
0 & \displaystyle -1/a & 0\vspace{0.01cm}\\
0 & 0 & \displaystyle 1/b\vspace{0.01cm}\\
 0 & 0 & \displaystyle -1/b
\end{array}
\right] \overline{s}
\,\leq \left[
\begin{array}{c}
1\\
1\\
1\\
1\\
1\\
1
\end{array}
\right]\!,
$$
$\overline{s}\, = (\Delta \subupscr{p}{L}{F}[1],\, \subupscr{p}{L}{F}[2],\, \subupscr{\beta}{L}{F})^T$, 
is symmetric with respect to the origin. The cones $\mathcal{C}_1^{\star},\ldots,\mathcal{C}_4^{\star}$ 
are given by
$$
\begin{array}{l}
\!\mathcal{C}_1^{\star} \!=\! \{g_1^T\, \overline{s} \leq 1 -
\frac{\subscr{V}{L}}{a},\, g_3^T\, \overline{s} \leq 1 -
\frac{\subscr{V}{L} \sin b}{a},\, g_5^T\, \overline{s} \leq 1 -
\frac{\subscr{\Omega}{L}}{b}\!\},\vspace{0.45cm}\\
\!\mathcal{C}_2^{\star} \!=\! \{g_1^T\, \overline{s} \leq 1 -
\frac{\subscr{V}{L}}{a},\, g_3^T\, \overline{s} \leq 1 -
\frac{\subscr{V}{L} \sin b}{a},\, g_6^T\,
\overline{s} \leq 1 - \frac{\subscr{\Omega}{L}}{b}\!\},\vspace{0.45cm}\\
\!\mathcal{C}_3^{\star} \!=\! \{g_1^T\, \overline{s} \leq 1 -
\frac{\subscr{V}{L}}{a},\, g_4^T\, \overline{s} \leq 1 -
\frac{\subscr{V}{L} \sin b}{a},\, g_5^T\, \overline{s} \leq 1 -
\frac{\subscr{\Omega}{L}}{b}\!\},\vspace{0.45cm}\\
\!\mathcal{C}_4^{\star} \!=\! \{g_1^T\, \overline{s} \leq 1 -
\frac{\subscr{V}{L}}{a},\, g_4^T\, \overline{s} \leq 1 -
\frac{\subscr{V}{L} \sin b}{a},\, g_6^T\, \overline{s} \leq 1 -
\frac{\subscr{\Omega}{L}}{b}\!\}.
\end{array}
$$
Condition (\ref{EQ:cond1_proof}) can then be rewritten as:
\vspace{-0.05cm}
\scriptsize
\begin{equation}\label{EQ:cond_proof_addit}
\begin{array}{l}
\!\!\left[
\begin{array}{l}
a(1-k_{11} + q_4 k_{21} -k_{12} + q_4 k_{22}) + b(q_2 - k_{13} + q_4 k_{23})\vspace{0.08cm}\\
a(1 - (k_{21}+k_{22})(d + q_3)) + b(1 + q_1 - k_{23}(d + q_3))\vspace{0.08cm}\\
- a (k_{21} + k_{22}) + b(1 - k_{23})
\end{array}
\right]  \!\in \mathcal{C}_1^{\star},\vspace{0.2cm}\\
\!\!\left[
\begin{array}{l}
a(1-k_{11} + q_4 k_{21} -k_{12} + q_4 k_{22}) - b(q_2 - k_{13} + q_4 k_{23}) \vspace{0.08cm}\\
a(1 - (k_{21}+k_{22})(d + q_3)) - b(1 + q_1 - k_{23}(d + q_3)) \vspace{0.08cm}\\
- a (k_{21} + k_{22}) - b(1 - k_{23})
\end{array}
\right] \!\in \mathcal{C}_2^{\star},\vspace{0.2cm}\\
\!\!\left[
\begin{array}{l}
a(1-k_{11} + q_4 k_{21} + k_{12} - q_4 k_{22}) + b(q_2 - k_{13} + q_4 k_{23}) \vspace{0.08cm}\\
a( -1 + (-k_{21} + k_{22})(d + q_3)) + b(1 + q_1 - k_{23}(d + q_3)) \vspace{0.08cm}\\
- a (k_{21} - k_{22}) + b(1 - k_{23})
\end{array}
\right] \!\in \mathcal{C}_3^{\star},\vspace{0.2cm}\\
\!\!\left[
\begin{array}{l}
a(1-k_{11} + q_4 k_{21} + k_{12} - q_4 k_{22}) - b(q_2 - k_{13} + q_4 k_{23}) \vspace{0.08cm}\\
a (-1 + (-k_{21} + k_{22})(d + q_3)) - b(1 + q_1 - k_{23}(d + q_3)) \vspace{0.08cm}\\
- a (k_{21} - k_{22}) - b(1 - k_{23})
\end{array}
\right]\! \in \mathcal{C}_4^{\star}.
\end{array}
\end{equation}
\normalsize
Because of the special structure of $B(q)$ in (\ref{Eq:AqBqEq}), we can select a
simplified state feedback matrix $K$ of the form~(\ref{Eq:feed_mat_K}): 
this allows to the decouple the control inputs
$\subscr{v}{F}$ and $\subscr{\omega}{F}$ and visualize the
polytope $\mathcal{K} \subset \mathbb{R}^3$ of all the feasible
gain matrices. We can then rewrite (\ref{EQ:cond_proof_addit})
in the following simplified form:
\begin{eqnarray}\label{Eq:All_cond1}
-k_{11} + q_4 k_{22} + \frac{b}{a}\,q_4\,k_{23} &\leq& - \frac{b}{a}\,q_2 - \frac{\subscr{V}{L}}{a}\,,\nonumber\vspace{0.4cm}\\
-(d+q_3)\,k_{22} - \frac{b}{a}\,(d+q_3)\,k_{23} &\leq& -\frac{b}{a}\,(1+q_1) - \frac{\subscr{V}{L} \sin b}{a}\,,\nonumber\vspace{0.4cm}\\
-k_{11} + q_4\,k_{22} - \frac{b}{a}\,q_4\,k_{23} &\leq& \frac{b}{a}\,q_2 - \frac{\subscr{V}{L}}{a}\,,\nonumber\vspace{0.4cm}\\
-(d+q_3)\,k_{22} + \frac{b}{a}\,(d+q_3)\,k_{23} &\leq& \frac{b}{a}\,(1+q_1) - \frac{\subscr{V}{L} \sin b}{a}\,,\nonumber\vspace{0.4cm}\\
-\frac{a}{b}\, k_{22} - k_{23} &\leq& - \frac{\subscr{\Omega}{L}}{b}\,,\nonumber\vspace{0.4cm}\\
\frac{a}{b}\,k_{22} - k_{23} &\leq& - \frac{\subscr{\Omega}{L}}{b}\,,\nonumber\vspace{0.4cm}\\
-k_{11} - q_4\,k_{22} + \frac{b}{a}\,q_4\,k_{23} &\leq& - \frac{b}{a}\,q_2 - \frac{\subscr{V}{L}}{a}\,,\nonumber\vspace{0.4cm}\\
-k_{11} - q_4\,k_{22} - \frac{b}{a}\,q_4\,k_{23} &\leq& \frac{b}{a}\,q_2 - \frac{\subscr{V}{L}}{a}\,.
\end{eqnarray}
The admissibility condition~(\ref{EQ:admiss}) leads to the additional \mbox{constraints}
\begin{equation}\label{Eq:All_cond2}
\begin{array}{l}
\begin{array}{ll}
\displaystyle k_{11} \leq \frac{\subscr{V}{F}}{a}, \;\;&\;
\qquad\qquad\;\;\displaystyle k_{11} \geq -\frac{\subscr{V}{F}}{a},
\end{array}
\vspace{0.35cm}\\
\begin{array}{ll}
\displaystyle k_{22} + \frac{b}{a}\,k_{23} \leq
\frac{\subscr{\Omega}{F}}{a}, \;\;&\,\quad
\displaystyle k_{22} - \frac{b}{a}\,k_{23} \leq \frac{\subscr{\Omega}{F}}{a},
\end{array}
\vspace{0.35cm}\\
\begin{array}{ll}
\displaystyle k_{22} - \frac{b}{a}\,k_{23} \geq -
\frac{\subscr{\Omega}{F}}{a}, \;\;&\;
\displaystyle k_{22} + \frac{b}{a}\,k_{23} \geq - \frac{\subscr{\Omega}{F}}{a}.
\end{array}
\end{array}
\end{equation}
The Fourier-Motzkin elimination is a mathematical algorithm for eliminating variables 
from a system of linear inequalities. 
Elimination of unknown $k_{ij}$ from the system of inequalities, consists in creating another system of the same kind
but without $k_{ij}$, such that both systems have the same solutions over the remaining variables.
\mbox{If one removes} all variables from a system of inequalities with \emph{numerical coefficients}, then one obtains a system 
of constant inequalities, which can be trivially decided to be true or false. 
This procedure can then be used to easily check whether a given system admits solutions or~not 
(see Appendix A for more details).\\ 
Applying the Fourier-Motzkin elimination to the inequalities (\ref{Eq:All_cond1})-(\ref{Eq:All_cond2}) 
with the assumption that $d > a$ (in~order to fix the sign of the coefficients of $k_{22}$
and $k_{23}$ in the second and fourth inequality of
(\ref{Eq:All_cond1})) we end up with the following conditions on the
variables $a$, $b$, $d$, $\subscr{V}{F}$, $\subscr{V}{L}$,
$\subscr{\Omega}{F}$,~$\subscr{\Omega}{L}$ and~uncertain
parameters $q_1,\ldots,q_4$ (the~detailed passages are reported in Appendix B),
$$
\begin{array}{l}
\subscr{\Omega}{L} \leq \frac{b(1+q_1) - \subscr{V}{L} \sin b}{d+q_3}\,,\;\; \frac{b(1+q_1) + \subscr{V}{L} \sin b}{d+q_3} \leq \subscr{\Omega}{F}\,,\vspace{0.22cm}\\
\subscr{V}{F} \geq \subscr{V}{L}\big(1 + \frac{q_4 \sin b}{d+q_3}\big) \!+ b \big(q_2 + \frac{q_4 (1+q_1)}{d+q_3}\big),\;\; \text{for}\;\; q_4 > 0\,,\vspace{0.25cm}\\
\subscr{V}{F} \geq \subscr{V}{L}\big(1 + \frac{q_4 \sin b}{d+q_3}\big) \!- b \big(q_2 + \frac{q_4 (1+q_1)}{d+q_3}\big),\;\; \text{for}\;\; q_4 > 0\,,\vspace{0.25cm}\\
\subscr{V}{F} \geq \subscr{V}{L} + b\, q_2,\hspace{3.88cm} \text{for}\;\; q_4 = 0\,,\vspace{0.15cm}\\
\subscr{V}{F} \geq \subscr{V}{L}\big(1 - \frac{q_4 \sin b}{d+q_3}\big) \!+ b \big(q_2 + \frac{q_4 (1+q_1)}{d+q_3}\big),\;\; \text{for}\;\; q_4 < 0\,,\vspace{0.25cm}\\
\subscr{V}{F} \geq \subscr{V}{L}\big(1 - \frac{q_4 \sin b}{d+q_3}\big) \!- b \big(q_2 + \frac{q_4 (1+q_1)}{d+q_3}\big),\;\; \text{for}\;\; q_4 < 0.
\end{array}
$$
An appropriate selection of the parameters $q_1,\ldots,q_4$ on the
extremes of the intervals (\ref{Eq:inter_q}), $($i.e., $q_1 = \frac{\sin b}{b} - 1$, $q_3 = a$, for the first inequality, $q_1 = 0$, $q_3 = -a$, for the second and $q_1 = 0$, $q_2 = \frac{1 - \cos b}{b}$, $q_3 = -a$, $q_4 = a$, for the third$)$, leads to
(\ref{EQ:cond1}) and~(\ref{EQ:cond2}). 
\hfill$\blacksquare$

Some remarks are in order at this point.\\
The inequalities (\ref{EQ:cond1}) and (\ref{EQ:cond2}) 
(which are linear in $\subscr{V}{F}$, $\subscr{V}{L}$, 
$\subscr{\Omega}{F}$, $\subscr{\Omega}{L}$ and
nonlinear in $a$, $b$, $d$), specify the role played 
by each of the parameters introduced in Assumption~\ref{Assumpt},  
in the solvability of the VMP. In particular, they show 
how the bounds on the forward and angular velocity of 
the follower robot are affected by the size of the visibility 
set $\mathcal{S}$ and the velocity of the leader.\\
Note that conditions (\ref{EQ:cond1}) and (\ref{EQ:cond2}) are
\emph{necessary} and \emph{sufficient} for the linear uncertain
system~(\ref{EQ:nonlin_unc_param}). Note also that owing to (\ref{EQ:cond2}), we have 
$\subscr{\Omega}{F} \geq \subscr{\Omega}{L}$.
Once fixed the variables $a$, $b$, $d$, $\subscr{V}{F}$,
$\subscr{V}{L}$, $\subscr{\Omega}{F}$, $\subscr{\Omega}{L}$
according to (\ref{EQ:cond1}) and (\ref{EQ:cond2}), the polytope
of all the feasible state feedback matrices is
given by~(\ref{Eq:All_cond1})-(\ref{Eq:All_cond2}): by
evaluating \mbox{(\ref{Eq:All_cond1})-(\ref{Eq:All_cond2})} on the 
vertices of the polyhedron~$\mathcal{Q}$, we can see that
$\mathcal{K}$ is defined by a set of 392 inequalities, most of whom 
are redundant (see, e.g., the external (green) 
polytope in~Fig.~\ref{FIG:plot_K}, below).
%

\begin{remark}[Selection of the gain matrix $K$]
Since the polytope $\mathcal{K}$ contains infinite gain matrices,
one needs an optimal criterion to select $K$, such as,~e.g.,
minimizing any matrix norm. In~the simulation experiments reported in
Sect.~\ref{SEC:simul}, we have chosen the matrix
$$
K =
\left[\begin{matrix}k_{11} & 0 & 0\vspace{0.05cm}\\
0 & k_{22} & k_{23}\end{matrix}\right],
$$ 
with minimum 2-norm.
\end{remark}


\vspace{-0.04cm}
\section{Extensions and applications}\label{SEC:Extensions}

In this section, we propose various extensions of the basic setup considered 
in Theorem~\ref{Teo1}, and discuss a few applications. We study the VMP in the presence 
of unknown but bounded disturbances, and consider the case of the leader moving 
along a circular path around a stationary target. We also extend
our results to chains of~robots.

\subsection{Rejection of unknown but bounded disturbances}\label{Subsect:LatDist}
\vspace{-0.2cm}

\noindent Let us consider the following system
\begin{equation}\label{EQ:nonlin_UBB}
\begin{array}{l}
\!\!\!\!\left[
\begin{matrix}\Delta \subupscr{\dot{p}}{L}{F}[1]\vspace{0.1cm}\\
\subupscr{\dot{p}}{L}{F}[2]\vspace{0.1cm}\\
\dot{\subupscr{\beta}{L}{F}}
\end{matrix} \right] = \left[\,
\begin{matrix}
0 \,&\, 0 & \frac{\cos\subupscr{\beta}{L}{F} - 1}{\subupscr{\beta}{L}{F}}\vspace{0.15cm}\\
0 \,&\, 0 & \frac{\sin\subupscr{\beta}{L}{F}}{\subupscr{\beta}{L}{F}}\vspace{0.15cm}\\
0 \,&\, 0 & 0
\end{matrix}\right]\!\!
\left[
\begin{matrix}
\Delta \subupscr{p}{L}{F}[1]\vspace{0.15cm}\\
\subupscr{p}{L}{F}[2]\vspace{0.15cm}\\
\subupscr{\beta}{L}{F}
\end{matrix} \right] +\vspace{0.3cm}\\ 
\hspace{-0.25cm}
\left[
\begin{matrix}
-1 \!&\! 0 \!&\! \subupscr{p}{L}{F}[2]\vspace{0.15cm}\\
0 \!&\! -1 \!&\! -\Delta \subupscr{p}{L}{F}[1] - d\vspace{0.15cm}\\
0 \!&\! 0 \!&\! -1
\end{matrix} \right]\!\!
\left[
\begin{matrix}
\subscr{v}{F}\vspace{0.15cm}\\
\subscr{h}{F}\vspace{0.15cm}\\
\subscr{\omega}{F}
\end{matrix} \right] +
\left[
\begin{matrix}
\cos\subupscr{\beta}{L}{F} \!&\! -\sin\subupscr{\beta}{L}{F} \!&\! 0\vspace{0.15cm}\\
\sin\subupscr{\beta}{L}{F} \!&\! \cos\subupscr{\beta}{L}{F} \!&\! 0\vspace{0.15cm}\\
0 \!&\! 0 \!&\! 1
\end{matrix}\right]\!\!
\left[
\begin{matrix}
\subscr{v}{L}\vspace{0.15cm}\\
\subscr{h}{L}\vspace{0.15cm}\\
\subscr{\omega}{L}
\end{matrix}\right]\!\!.
\end{array}
\end{equation}
\normalsize With respect to (\ref{EQ:nonlin_unc_param}),
two new components, $\subscr{h}{F}$ and $\subscr{h}{L}$, are
present in the input and disturbance vectors $u$ and~$\delta$. They are \emph{unknown
but bounded} disturbances acting on the robots $\textup{F}$ and~$\textup{L}$ 
(e.g., lateral wind in a real setup). Our purpose here is to solve the VMP in 
the presence of the disturbances $\subscr{h}{F}$,~$\subscr{h}{L}$.\\
Collecting together all the perturbations acting on the nominal
system (i.e., $\subscr{v}{L}$, $\subscr{\omega}{L}$,
$\subscr{h}{F}$ and~$\subscr{h}{L}$), we~can~rewrite~(\ref{EQ:nonlin_UBB}) as
\begin{equation}\label{EQ:nonlin_UBB2}
\begin{array}{l}
\!\left[
\begin{matrix}\Delta \subupscr{\dot{p}}{L}{F}[1]\vspace{0.15cm}\\
\subupscr{\dot{p}}{L}{F}[2]\vspace{0.15cm}\\
\dot{\subupscr{\beta}{L}{F}}
\end{matrix} \right] = \left[\,
\begin{matrix}
0 \,&\, 0 & \frac{\cos\subupscr{\beta}{L}{F} - 1}{\subupscr{\beta}{L}{F}}\vspace{0.15cm}\\
0 \,&\, 0 & \frac{\sin\subupscr{\beta}{L}{F}}{\subupscr{\beta}{L}{F}}\vspace{0.15cm}\\
0 \,&\, 0 & 0
\end{matrix}\right]\!\!\!
\left[
\begin{matrix}
\Delta \subupscr{p}{L}{F}[1]\vspace{0.15cm}\\
\subupscr{p}{L}{F}[2]\vspace{0.15cm}\\
\subupscr{\beta}{L}{F}
\end{matrix} \right]\, + \vspace{0.25cm}\\ 
\!\left[
\begin{matrix}
-1 \!&\! \subupscr{p}{L}{F}[2]\vspace{0.15cm}\\
0 \!&\! -\Delta \subupscr{p}{L}{F}[1] - d\vspace{0.15cm}\\
0 \!&\! -1
\end{matrix} \right]\!\!\!
\left[
\begin{matrix}
\subscr{v}{F}\vspace{0.15cm}\\
\subscr{\omega}{F}
\end{matrix}\right] \!\!+\!\!
\left[
\begin{array}{cc|cc}
\cos\subupscr{\beta}{L}{F} \!&\! 0 \,& 0 &\! -\sin\subupscr{\beta}{L}{F}\!\vspace{0.02cm}\\
\sin\subupscr{\beta}{L}{F} \!&\! 0 \,& -1 &\! \cos\subupscr{\beta}{L}{F}\vspace{0.02cm}\\
0 & 1 \,& 0 \!&\! 0
\end{array}
\right]\!\!\!
\left[\!
\begin{MAT}[3pt]{c}
\subscr{v}{L}\\
\subscr{\omega}{L}\\
\subscr{h}{F}\\
\subscr{h}{L}
\addpath{(0,2,0)l}\\
\end{MAT}\!\right]\!\!.
\end{array}
\end{equation}
\begin{figure}[t!]
\begin{center}
\psfrag{a}{$k_{11}$}\psfrag{b}{$k_{22}$}\psfrag{c}{$k_{23}$}
\includegraphics[width=0.92\columnwidth]{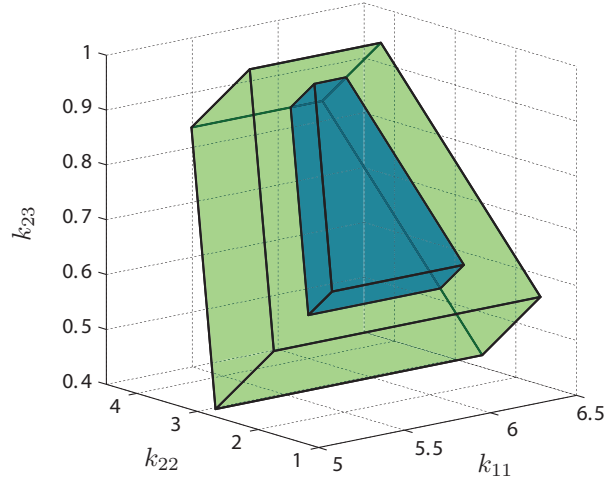}
\caption{\textit{Polytopes $\mathcal{K}$ for a set of given
parameters:} (\emph{blue}, \mbox{internal}) with disturbances; (\emph{green}, external)
without \mbox{disturbances}.}\label{FIG:plot_K}
\end{center}
\end{figure}
\noindent Let $\mathcal{U}$ be given in~(\ref{EQ:boundsUD}), and
define
\begin{equation}\label{EQ:D-dist}
\begin{array}{c}
\!\!\!\mathcal{D} =\! \{(\subscr{v}{L}, \subscr{\omega}{L}, \subscr{h}{F},
\subscr{h}{L})^T :\, \subscr{v}{L} \in [-\subscr{V}{L},\,\subscr{V}{L}],\vspace{0.2cm}\\ 
\subscr{\omega}{L} \in [-\subscr{\Omega}{L},\,\subscr{\Omega}{L}],\,
\subscr{h}{F} \in [-\subscr{H}{F},\,\subscr{H}{F}],\, 
\subscr{h}{L} \in [-\subscr{H}{L},\,\subscr{H}{L}]\},
\end{array}
\end{equation}
where $\subscr{H}{F}$, $\subscr{H}{L}$ are strictly positive
constants. Using the same arguments as those in Theorem~\ref{Teo1}, 
we can prove the following corollary (note that the feedback matrix
$K$ is again of the form~(\ref{Eq:feed_mat_K})).
\begin{corollary}[VMP with disturbances]\label{Cor_UBB} 
Choose $\mathcal{U}$ and $\mathcal{S}$ as in Assumption~\ref{Assumpt}, 
$\mathcal{D}$ as in (\ref{EQ:D-dist}), and
let $d > a$, $0 < b \leq \pi/2$. The VMP for the robots $\textup{F}$ and $\textup{L}$
in the presence of the unknown but bounded disturbances $\subscr{h}{F}$,
$\subscr{h}{L}$, has a solution if the following conditions are
satisfied
\begin{equation}\label{EQ:condUBB1}
\begin{array}{l}
\hspace{-0.1cm}\displaystyle\subscr{V}{F} \geq \subscr{V}{L}\left(1 +
\frac{a\sin b}{d-a}\right) + 1 - \cos b\vspace{0.2cm}\\
\hspace{2.5cm}\displaystyle +\; \frac{a(\subscr{H}{F} + \subscr{H}{L} +\, b)}{d-a} + \subscr{H}{L} \sin b,
\end{array}
\end{equation}
\vspace{0.1cm}
\begin{equation}\label{EQ:condUBB2}
\begin{array}{l}
\hspace{-2.7cm}\displaystyle\subscr{\Omega}{L} \leq \frac{(1-\subscr{V}{L})\sin
b\, - (\subscr{H}{F}\, + \subscr{H}{L})}{d+a}\,,\vspace{0.5cm}\\
\hspace{-2.7cm}\displaystyle \subscr{\Omega}{F} \geq \frac{\subscr{V}{L}\sin b + b + (\subscr{H}{F}\, +
\subscr{H}{L})}{d-a}.
\end{array}
\end{equation}
\end{corollary}
\vspace{0.1cm}

\begin{remark}
Note that because of the additional terms
$\subscr{H}{F}$ and $\subscr{H}{L}$, conditions
(\ref{EQ:condUBB1})-(\ref{EQ:condUBB2}) are stricter than
(\ref{EQ:cond1})-(\ref{EQ:cond2}) and then the polytope
$\mathcal{K}$ is smaller in this case. This is evident in
Fig.~\ref{FIG:plot_K}, where the polytope $\mathcal{K}$ (\emph{blue}, internal)
obtained for $a=0.15$~m, $b=\pi/3$ rad, $d=1.6$ m,
$\subscr{V}{F}=0.95$~m/s, $\subscr{V}{L}=0.1$ m/s,
$\subscr{\Omega}{F} = \pi/2$ rad/s, $\subscr{\Omega}{L} = \pi/20$
rad/s and $\subscr{H}{F} = 0.2$~m/s, $\subscr{H}{L} = 0.1$~m/s is
compared with the polytope (\emph{green}, external) corresponding to $\subscr{H}{L} =
\subscr{H}{F} = 0$ m/s.
\end{remark}

\vspace{-0.1cm}
\subsection{VMP on a circle}\label{Subsect:AngPar}
\vspace{-0.2cm}

\begin{figure}[t!]
\begin{center}
\psfrag{a}{$\gamma$} \psfrag{b}{$\rho$}
\psfrag{c}{$\textup{F}$} \psfrag{d}{$\textup{L}$}
\psfrag{e}{$\large\mathcal{S}$}
\psfrag{f}{\!\!\!\!\!\!\!\!\!\large{$\big(\frac{\sin\gamma}{\rho},\,\frac{1-\cos\gamma}{\rho}\big)$}}
\psfrag{g}{$\subscr{x}{L}$}
\psfrag{h}{$\subscr{y}{L}$}
\includegraphics[width=0.75\columnwidth]{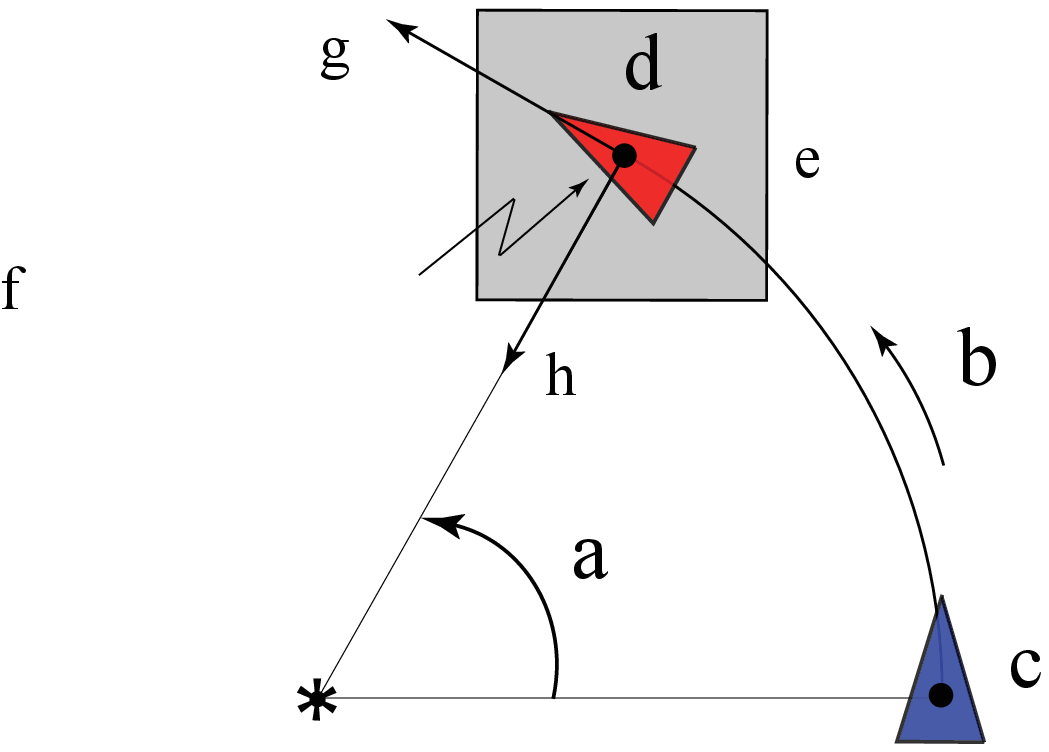}
\caption{\textit{VMP on a circle:} pose of the robots $\textup{L}$ and~$\textup{F}$ for
$(\Delta \subupscr{p}{L}{F}[1], \Delta \subupscr{p}{L}{F}[2],
\Delta \subupscr{\beta}{L}{F})^T = (0,\,0,\,0)^T$, $\subscr{v}{L} = \subscr{v}{F}$ and 
$\subscr{\omega}{L} = \subscr{\omega}{F} = \rho$.}\label{FIG:ring}
\end{center}
\end{figure}

In this section, we will suppose that the leader robot moves along a circular path, around a static target.
This scenario could be of interest in several real-world applications, such as, e.g., for
environmental surveillance, patrolling or terrain and utilities 
inspection~\cite{CasbeerKiBeMcLiMe_IJSC06,SuscaBuMa_TCST08}.
Differently from Sect.~\ref{SUBSEC:Sol}, we will assume that the pose 
of robot $\textup{L}$ with respect to the frame of $\textup{F}$ is defined through 
the angle $0 < \gamma < \pi/2$ and the angular velocity $\rho > 0$,
instead of the distance parameter~$d$ (see~Fig.~\ref{FIG:ring}).
Let us consider the following change of variables in
system~(\ref{EQ:nonlin_dubin}):
$$
\begin{array}{c}
(\subupscr{p}{L}{F}[1], \subupscr{p}{L}{F}[2], \subupscr{\beta}{L}{F})^T
\rightarrow\, (\Delta \subupscr{p}{L}{F}[1], \Delta
\subupscr{p}{L}{F}[2], \Delta \subupscr{\beta}{L}{F})^T,\vspace{0.2cm}\\
\hspace{-0.82cm} (\subscr{v}{L},\subscr{\omega}{L})^T \rightarrow\, (\subscr{v}{L},\,
\subscr{\omega}{L})^T,\vspace{0.2cm}\\
\hspace{-0.55cm} (\subscr{v}{F},\subscr{\omega}{F})^T \rightarrow\, (\subscr{v}{F},\,
\Delta \subscr{\omega}{F})^T,
\end{array}
$$
where 
$\Delta \subupscr{p}{L}{F}[1] = \subupscr{p}{L}{F}[1] -
\frac{\sin \gamma}{\rho}$, $\Delta \subupscr{p}{L}{F}[2] =
\subupscr{p}{L}{F}[2] - \frac{1 -\cos \gamma}{\rho}$, 
$\Delta\subupscr{\beta}{L}{F} = \subupscr{\beta}{L}{F} -\gamma$
and
$\Delta \subscr{\omega}{L} =
\subscr{\omega}{L} - \rho$, $\Delta \subscr{\omega}{F} =
\subscr{\omega}{F} - \rho$. 
Following the same procedure detailed in the previous sections, we obtain the system\vspace{-0.2cm}
\begin{equation}\label{EQ_sysring}
\begin{array}{l}
\hspace{-0.15cm}\left[
\begin{matrix}\Delta\subupscr{\dot{p}}{L}{F}[1]\vspace{0.1cm}\\
\Delta\subupscr{\dot{p}}{L}{F}[2]\vspace{0.1cm}\\
\Delta\dot{\subupscr{\beta}{L}{F}}
\end{matrix} \right] = \left[\begin{matrix}
0 \,&\; \rho \,&\; q_2-\sin\gamma\vspace{0.1cm}\\
-\rho \,&\, 0 \,&\; q_1+\cos\gamma\vspace{0.1cm}\\
0 \,&\; 0 \,&\; 0
\end{matrix} \right]\!\!
\left[
\begin{matrix}
\Delta \subupscr{p}{L}{F}[1]\vspace{0.1cm}\\
\Delta \subupscr{p}{L}{F}[2]\vspace{0.1cm}\\
\Delta \subupscr{\beta}{L}{F}
\end{matrix} \right]\vspace{0.32cm}\\ 
\hspace{-0.05cm}+ \!\left[\begin{matrix}
-1 &\;\, q_4+\frac{1 - \cos\gamma}{\rho}\vspace{0.1cm}\\
0 &\;\, -q_3-\frac{\sin\gamma}{\rho}\vspace{0.1cm}\\
0 &\;\, -1
\end{matrix} \right]\!\!
\left[
\begin{matrix}
\subscr{v}{F}\vspace{0.1cm}\\
\Delta\subscr{\omega}{F}
\end{matrix} \right] + \left[\begin{matrix}
q_5 + \cos\gamma & 0\vspace{0.1cm}\\
q_6 + \sin\gamma & 0\vspace{0.1cm}\\
0 & 1
\end{matrix} \right]\!\!
\left[
\begin{matrix}
\subscr{v}{L}\vspace{0.1cm}\\
\Delta\subscr{\omega}{L}
\end{matrix}\right]\!,
\end{array}
\end{equation}
where
$$
\begin{array}{ll}
\displaystyle q_1 &= \frac{\sin(\Delta\subupscr{\beta}{L}{F} + \gamma) - \sin\gamma}{\Delta \subupscr{\beta}{L}{F}}
- \cos\gamma, \vspace{0.3cm}\\ 
\displaystyle q_2 &= \frac{\cos(\Delta\subupscr{\beta}{L}{F} + \gamma) -
\cos\gamma}{\Delta \subupscr{\beta}{L}{F}} +
\sin\gamma,
\end{array}
$$
$$
\begin{array}{ll}
\displaystyle q_3 &= \Delta \subupscr{p}{L}{F}[1],\;\; q_5 =
\cos(\Delta\subupscr{\beta}{L}{F} + \gamma) - \cos\gamma, \vspace{0.22cm}\\
q_4 &= \Delta \subupscr{p}{L}{F}[2],\;\; q_6 = \sin(\Delta\subupscr{\beta}{L}{F} + \gamma) - \sin\gamma.
\end{array}
$$
\vspace{-0.4cm}
\begin{assumption}\label{Assumpt2}
Let us suppose that
\begin{equation*}\label{EQ:boundsUD2}
\begin{array}{l}
\hspace{-0.07cm}\mathcal{U} = \!\left\{(\subscr{v}{F},
\Delta\subscr{\omega}{F})^T\!: 
\subscr{v}{F} \in [- \subscr{V}{F},\,\subscr{V}{F}],\, \Delta\subscr{\omega}{F} 
\in [-\subscr{\Omega}{F},\,\subscr{\Omega}{F}]\right\}\!,\vspace{0.2cm}\\
\hspace{-0.07cm}\mathcal{D} = \!\left\{(\subscr{v}{L},
\Delta\subscr{\omega}{L})^T\!: \subscr{v}{L} \in
[-\subscr{V}{L},\,\subscr{V}{L}],\, \Delta\subscr{\omega}{L} \in
[-\subscr{\Omega}{L},\,\subscr{\Omega}{L}]\right\}\!,
\end{array}
\end{equation*}
where $0 < \subscr{V}{F} < 1$, $0 < \subscr{V}{L} < 1$ and
$0 < \subscr{\Omega}{F}$, $0 < \subscr{\Omega}{L} <~\rho$. Let us also 
consider the following visibility~set (see Fig.~\ref{FIG:ring})
\begin{equation}\label{EQ:VisBox2}
\begin{array}{c}
\!\!\!\mathcal{S} = \{(\Delta \subupscr{p}{L}{F}[1],
\Delta \subupscr{p}{L}{F}[2], \Delta\subupscr{\beta}{L}{F})^T\!:\, \Delta
\subupscr{p}{L}{F}[1] \in [-a,\,a],\vspace{0.15cm}\\ 
\;\;\; \Delta\subupscr{p}{L}{F}[2] \in
[-a,\,a],\; \Delta\subupscr{\beta}{L}{F} \in [-b,\,b]\},
\end{array}
\end{equation}
where $a > 0$ and $b > 0$.
\end{assumption}
Since the state of system (\ref{EQ_sysring}) is 
constrained in (\ref{EQ:VisBox2}), the polyhedron
$\mathcal{Q} \subset \mathbb{R}^6$ is defined as follows:
$$
\begin{array}{l}
\!q_1 \!\in\! \big[\,\frac{\sin(b + \gamma) - \sin\gamma}{b} - \cos\gamma, \frac{\sin(b - \gamma) + \sin\gamma}{b} - \cos\gamma\big],\vspace{0.15cm}\\ 
\!q_2 \!\in\! [\frac{\cos(b + \gamma) - \cos\gamma}{b} + \sin\gamma, \frac{-\cos(b - \gamma) + \cos\gamma}{b} + \sin\gamma],\vspace{0.15cm}\\ 
\!q_3 \!\in\! [-a,\, a],\, q_5 \!\in\! [\cos(b+\gamma) - \cos\gamma, \cos(b-\gamma) - \cos\gamma], \vspace{0.15cm}\\ 
\!q_4 \!\in\! [-a,\, a],\, q_6 \!\in\! [-\!\sin(b-\gamma) - \sin\gamma, \sin(b+\gamma) - \sin\gamma].
\end{array}
$$
The proof of the next theorem is analogous to that of
Theorem~\ref{Teo1} and it is omitted. The feedback matrix $K$
has also in this case the form~(\ref{Eq:feed_mat_K}).

\begin{theorem}[Solvability of the VMP on a circle]\label{Theo_circ}
Choose $\mathcal{U}$, $\mathcal{D}$ and $\mathcal{S}$ as in Assumption~\ref{Assumpt2}
and let \mbox{$1-\cos\gamma > \rho a$}, $0 \leq b \pm \gamma \leq \pi/2$. 
The~VMP on a circle has a solution if the following conditions are satisfied 
$$
\begin{array}{l}
\subscr{V}{F} \geq \subscr{V}{L}\,\big(\!\cos(b-\gamma) +
\frac{\sin(b+\gamma)(1-\cos\gamma - \rho\,a)}{\sin\gamma+\rho a}\big) \!+ \cos\gamma + \rho a\vspace{0.28cm}\\ \hspace{0.26cm}-\frac{1-\cos\gamma - \rho\,a}{\sin\gamma + \rho a}\,\left(\sin(b+\gamma) - \sin\gamma + \rho a\right) - \cos(b+\gamma),
\end{array}
$$
\vspace{1mm}
\begin{equation}\label{Eq:Omega_circ}
\begin{array}{l}
\hspace{-0.425cm}\displaystyle\subscr{\Omega}{L} \leq\, \rho\left(\frac{(1-\subscr{V}{L})\sin (b+\gamma)}{\sin\gamma+\rho a} - 1\right)\!,\vspace{0.6cm}\\
\hspace{-0.425cm}\displaystyle\subscr{\Omega}{F} \geq\,
\rho\left(\frac{\subscr{V}{L}\,\sin(b+\gamma) + \sin (b-\gamma) +
\sin\gamma + \rho a}{\sin\gamma - \rho a}\right)\!.
\end{array}
\end{equation}
\end{theorem}
Note that differently from Theorem~\ref{Teo1}, in this case, \mbox{owing} to (\ref{Eq:Omega_circ})
is \emph{not always} true (i.e., for all values of the parameters) 
that $\subscr{\Omega}{F} \geq \subscr{\Omega}{L}$.

\vspace{0.1cm}
\subsection{Chain of robots}\label{SEC:ext_chain}
\vspace{-0.1cm}

\begin{figure}[t!]
\begin{center}
\psfrag{a}{\!\!$\mathcal{S}_2$}
\psfrag{b}{\!\!\!$\mathcal{S}_3$}
\psfrag{c}{\!\!\!$\mathcal{S}_{n}$}
\psfrag{e}{1}
\psfrag{f}{2}
\psfrag{g}{$n-1$}
\psfrag{h}{$n$}
\includegraphics[width=0.94\columnwidth]{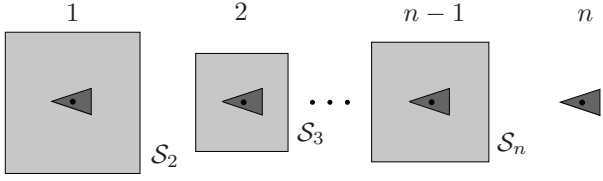}
\vspace{0.03cm}
\caption{Chain of $n$ robots.}\label{FIG:Chain}
\end{center}
\end{figure}

\begin{figure}[b!]
\begin{center}
\psfrag{a}{$V_k$} 
\psfrag{b}{$\Omega_k$}
\psfrag{c}{$b_k$} 
\psfrag{d}{$k$}
\includegraphics[width=0.8\columnwidth]{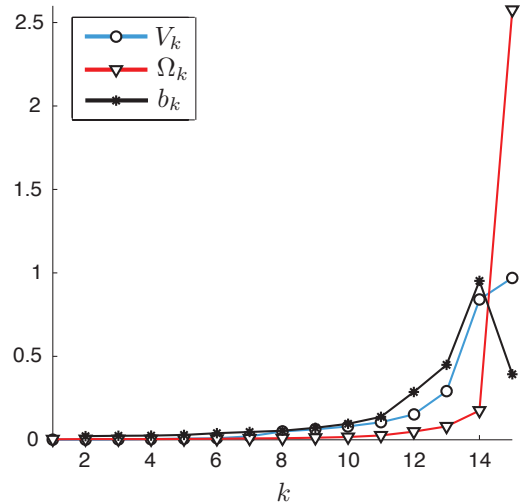}
\vspace{-0.2cm}
\caption{Progression of a feasible set of parameters $V_k$, $\Omega_k$, $b_k$ 
for a chain of $n = 15$ robots ($a_k = 0.1$, $d_k = 7$, for all $k$). Angles are in radians.}\label{FIG:Evol_par}
\end{center}
\end{figure}

Next, we consider more complex robotic networks built by concatenating multiple 
leader-follower units (see~Fig.~\ref{FIG:Chain}). When equipped with wireless sensors, these arrays of robots
could be used, for example, to efficiently monitor the temperature and/or salinity 
of the ocean or measure the average concentration of air pollutants~\cite{LynchScYaFr_TRO08}. 
In what follows, the feasibility conditions of Theorem~\ref{Teo1} will be extended to such 
robot chains in order to maintain network-wide visibility between the agents.\\
Because of the leader-follower hierarchy within the chain,
vehicle $k+1$ (the~``follower'') will aim at keeping the vehicle ahead (robot~$k$, the ``leader''), 
in its visibility set. Let $a_k$, $b_k$, $d_k$, $d_k > a_k$, \mbox{$b_k \leq \pi/2$}, $k \in \{2,\ldots,n\}$, $n > 2$, 
be the strictly positive parameters defining the visibility set $\mathcal{S}_k$ of robot $k$-th and let \mbox{$0 < V_k < 1$}, 
$\Omega_k > 0$, $k \in \{1,\ldots,n\}$ be the bounds on $k$-th robot's linear and angular velocities (recall Assumption~\ref{Assumpt}). By~propagating conditions (\ref{EQ:cond1})-(\ref{EQ:cond2}) of Theorem~\ref{Teo1} starting from robot~1 (that guides the 
formation), we obtain the following set of inequalities (linear in $V_k$ and~$\Omega_k$)
\begin{equation}\label{EQ:chain1}
\begin{array}{l}
\hspace{-0.4cm}\displaystyle V_{k+1} \geq V_{k}\left(1 + \frac{a_{k+1}\sin b_{k+1}}{d_{k+1}\,-\,a_{k+1}}\right)+ 1 - \cos b_{k+1}\vspace{0.36cm}\\
\displaystyle \hspace{0.435cm}+ \frac{a_{k+1}\, b_{k+1}}{d_{k+1}\,-\,a_{k+1}},\;\; k \in \{1,\ldots,n-1\},
\end{array}
\end{equation} 
\begin{equation}
\Omega_1 \leq \frac{(1 - V_1)\sin b_2}{d_2\,+\,a_2}\,,\;\;\,  \Omega_{n} \geq \frac{V_{n-1}\sin b_{n} + b_n}{d_{n}\,-\,a_{n}},\vspace{0.17cm}\label{EQ:chain2}
\end{equation} 
and for all $k \in \{2,\ldots,n - 1\}$,
\begin{equation}\label{EQ:chain3}
\begin{array}{c}
\displaystyle\frac{V_{k-1}\sin b_k + b_k}{d_k - a_k} \leq\, \Omega_k \,\leq
\frac{(1 - V_k)\sin b_{k+1}}{d_{k+1} +a_{k+1}}.
\end{array}
\end{equation} 
It is an easy task to verify that if $a_k = a$, $b_k = b$, $d_k = d$,
$k \in \{2,\ldots,n\}$ and $V_k = V$, $k \in \{1,\ldots,n\}$, 
($a > 0$, $d > a$, $0 < b \leq \pi/2$, $0 < V < 1$), 
condition (\ref{EQ:chain3}) (and condition (\ref{EQ:chain1}) as well) is \emph{not} satisfied. 
Nevertheless, it~can be proved that if $V_k$ and at least one of the
three parameters defining the visibility sets are left free to vary
from robot to robot, then (\ref{EQ:chain1})-(\ref{EQ:chain3}) can be always fulfilled.
Fig.~\ref{FIG:Evol_par} shows the progression of a feasible set of parameters $V_k$, $\Omega_k$, $b_k$ 
for a chain of $n = 15$ robots, when $a_k = a = 0.1$ and $d_k = d = 7$, $k \in \{2,\ldots,15\}$.
%
%
%
\begin{figure*}[t!]
\begin{center}
\begin{tabular}{ccc}
\!\!\!\!\!\subfigure[]{\includegraphics[width=0.61\columnwidth]{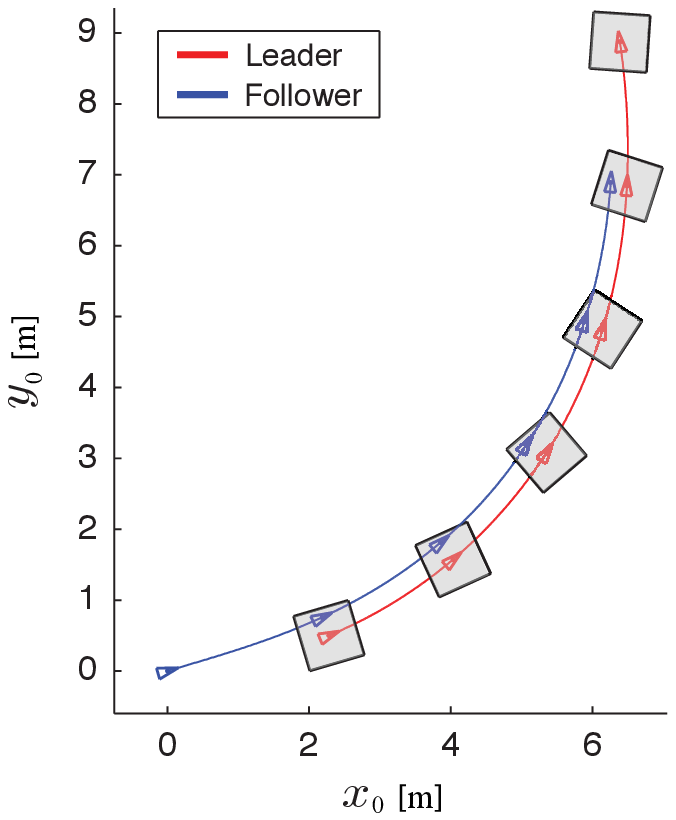}}
\,&
\psfrag{a}{\scriptsize{$\Delta \subupscr{p}{L}{F}[1]$\,
[m]}} \psfrag{b}{\scriptsize{$\subupscr{p}{L}{F}[2]$\, [m]}}
\psfrag{c}{\scriptsize{$\subupscr{\beta}{L}{F}$\, [rad]}}
\subfigure[]{\includegraphics[width=0.68\columnwidth]{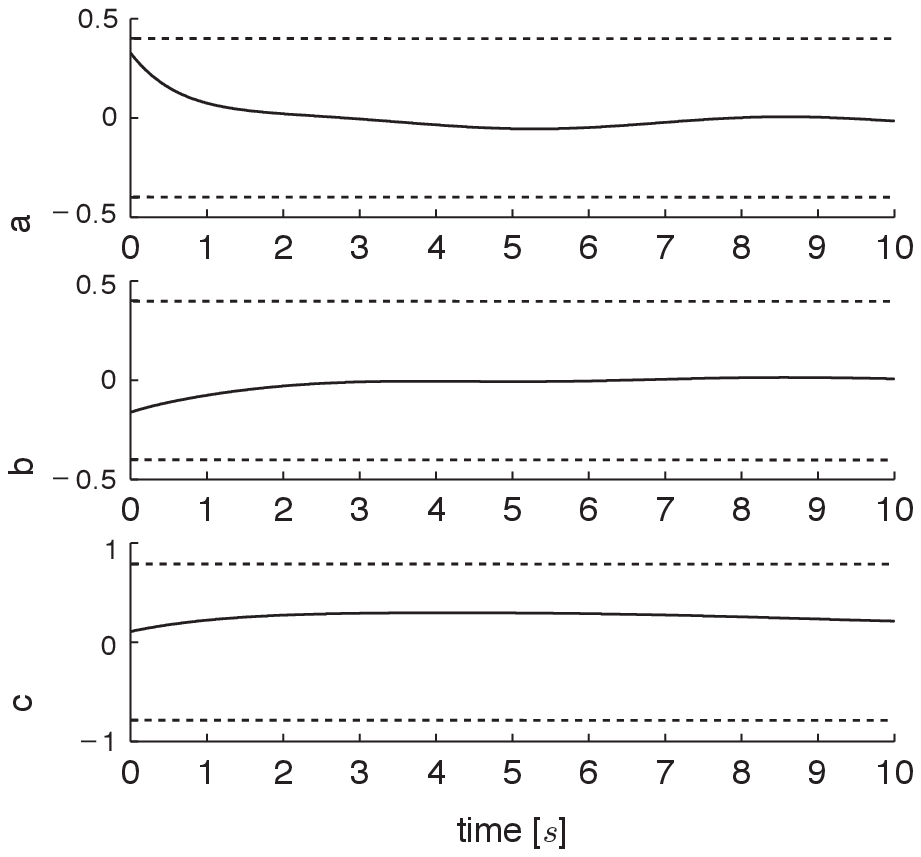}}
\,&
\psfrag{a}{\footnotesize{$\subscr{v}{F}$}\,
\scriptsize{[m/s]}}
\psfrag{b}{\footnotesize{$\subscr{\omega}{F}$}\,
\scriptsize{[rad/s]}}
\subfigure[]{\includegraphics[width=0.68\columnwidth]{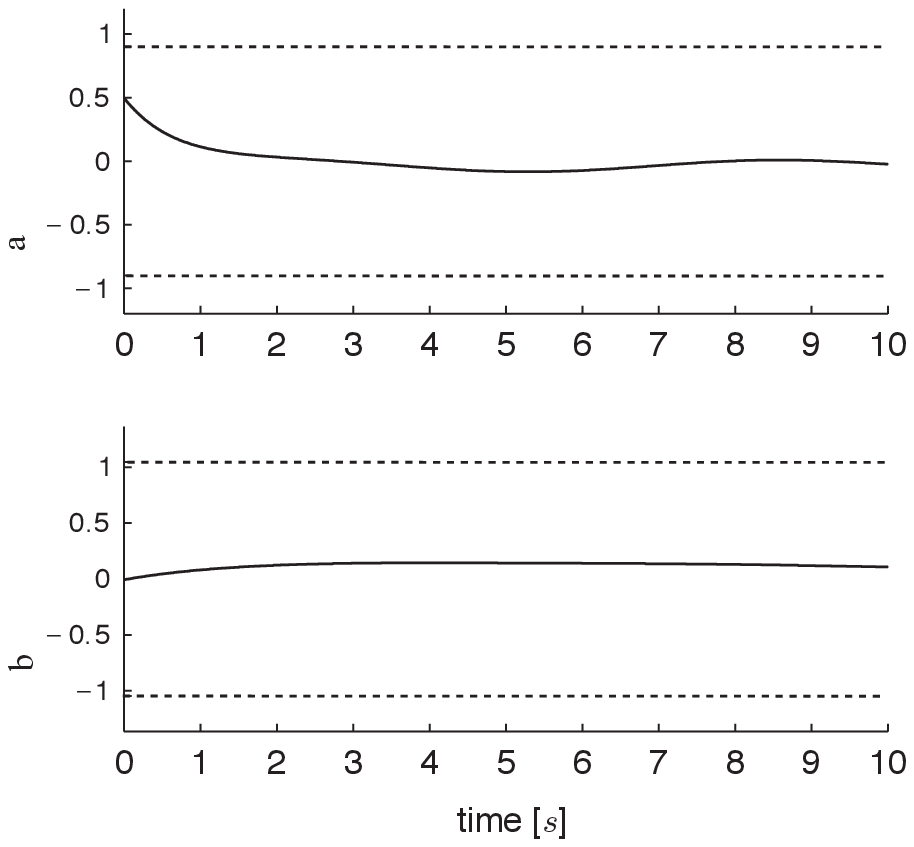}}
\end{tabular}
\end{center}
\caption{\emph{Basic scenario}: (a) Trajectory of the leader and follower, and
visibility set $\mathcal{S}$; (b) $\Delta\subupscr{p}{L}{F}[1]$, $\subupscr{p}{L}{F}[2]$, $\subupscr{\beta}{L}{F}$ (solid)
and bounds $\pm a$, $\pm a$, $\pm b$ (dash); (c)~$\subscr{v}{F}$, $\subscr{\omega}{F}$ (solid) and bounds \mbox{$\pm
\subscr{V}{F}$, $\pm \subscr{\Omega}{F}$ (dash)}.}\label{FIG:simulaz1}
\end{figure*}
We~conclude this section with the following corollary and a few remarks:
\begin{corollary}
The chain of robots can hold only a \emph{finite} number of agents and cannot be \emph{closed}, i.e., 
robot~$1$ cannot cyclically pursue robot $n$. 
\end{corollary}
\emph{Proof:}\,
Let us study the connection existing between parameters $V_1$ and~$V_n$. 
After simple algebraic manipulations on (\ref{EQ:chain1}), we obtain the following inequality: 
\begin{equation}\label{Eq:chain_proof_ineq}
\begin{array}{c}
\hspace{-2.5cm}\displaystyle V_1 \,\leq\,
\frac{1}{\displaystyle\prod\limits_{i=1}^n 
\Big(1 + \frac{a_i\,\sin b_i}{d_i - a_i}\Big)}\;V_n\vspace{0.05cm}\\ 
\hspace{2.5cm}\displaystyle -\,\sum_{i=2}^n \;\,\frac{\displaystyle 1 - \cos b_i + \frac{a_i\,\sin b_i}{d_i - a_i}}{\displaystyle\prod\limits_{k=2}^i \Big(1 + \frac{a_k\,\sin b_k}{d_k - a_k}\Big)}\,. \vspace{-0.15cm}
\end{array}
\end{equation}
When the number of robots $n$ tends towards infinity, the first term on the right-hand side 
of~(\ref{Eq:chain_proof_ineq}) converges to zero (for any choice of the parameters $a_i$, $b_i$, $d_i$ satisfying the previous assumptions) 
asymptotically leading to the inequality $V_1 \leq 0$, which contradicts the initial hypothesis of a strictly positive $V_1$. 
On the other hand, if the chain of robots were closed, the following additional 
inequality should be satisfied
$$
\displaystyle V_{1} \,\geq\, V_{n}\left(1 + \frac{a_{1}\sin b_{1}}{d_{1} - a_{1}}\right) +\, 1 \,-\, \cos b_{1} + \frac{a_{1}\, b_{1}}{d_{1} - a_{1}},
$$
but it is incompatible with condition~(\ref{Eq:chain_proof_ineq}). 
\hfill$\blacksquare$
\vspace{-0.25cm}

\indent It is easy to prove from condition~(\ref{Eq:chain_proof_ineq}), that once fixed a rule for the evolution of the parameters defining the visibility sets (i.e., $a_{k} = f_a(k)$, $b_{k} = f_b(k)$, $d_{k} = f_d(k)$, with $f_a,\,f_b,\,f_d: \mathbb{Z}_{> 1} \rightarrow \mathbb{R}$ given discrete maps), an upper bound on the maximum number of robots the chain can hold is given by the maximum positive integer $N$ that fulfills the following inequality:
$$
\sum_{i=2}^N \Big(1 - \cos b_i + \frac{a_i\,b_i}{d_i - a_i}\Big) \!\prod_{k\,=\,i+1}^N \Big(1 + \frac{a_k\,\sin b_k}{d_k - a_k}\Big) < 1.
$$
\;\; Note that once the state feedback matrices $K_i$, $i \in
\{2,\ldots,n\}$ of the robots have been established, the
implementation of the control laws is \emph{totally distributed}: 
in~fact, each agent only needs to know the relative position and
orientation of the preceding vehicle in the chain, 
to execute its control action.
%
%
%
%
%
\begin{figure*}[t!]
\begin{center}
\begin{tabular}{ccc}
\!\!\!\!\!\!\subfigure[]{\includegraphics[width=0.83\columnwidth]{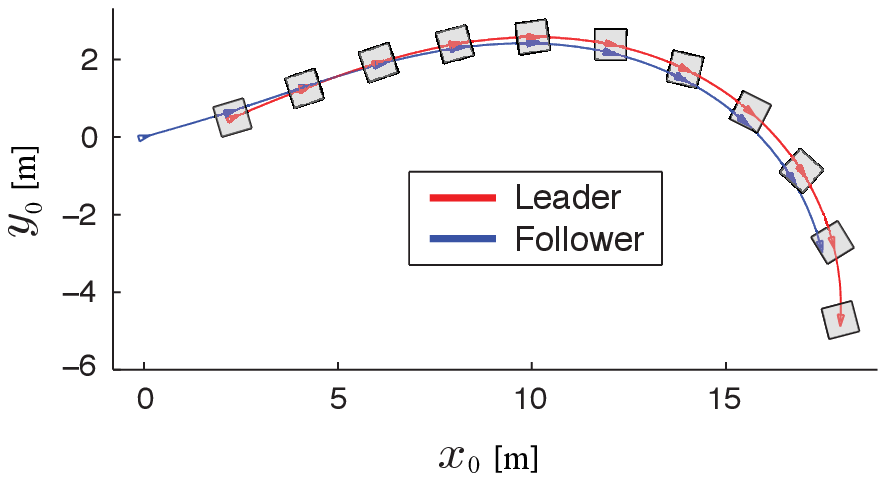}}
&
\psfrag{c}{\scriptsize{$\Delta \subupscr{p}{L}{F}[1]$\,
[m]}} \psfrag{b}{\scriptsize{$\subupscr{p}{L}{F}[2]$\, [m]}}
\psfrag{a}{\scriptsize{$\subupscr{\beta}{L}{F}$\, [rad]}}
\subfigure[]{\includegraphics[width=0.587\columnwidth]{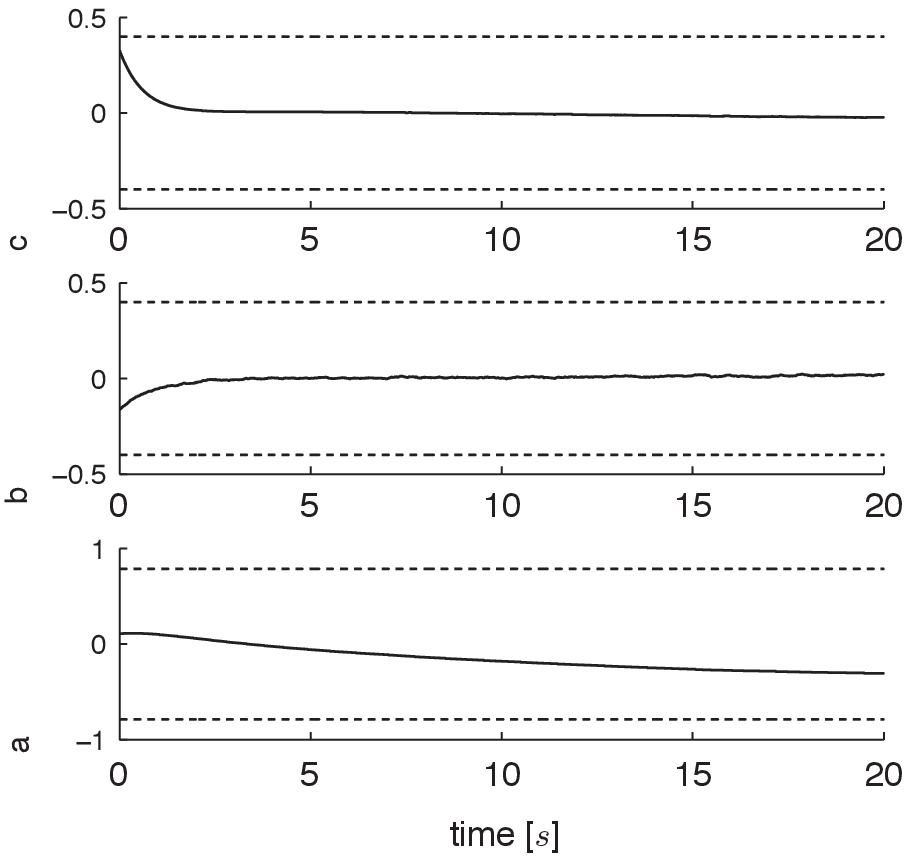}}
&
\psfrag{a}{\footnotesize{$\subscr{v}{F}$}\,
\scriptsize{[m/s]}}
\psfrag{b}{\footnotesize{$\subscr{\omega}{F}$}\,
\scriptsize{[rad/s]}}
\subfigure[]{\includegraphics[width=0.602\columnwidth]{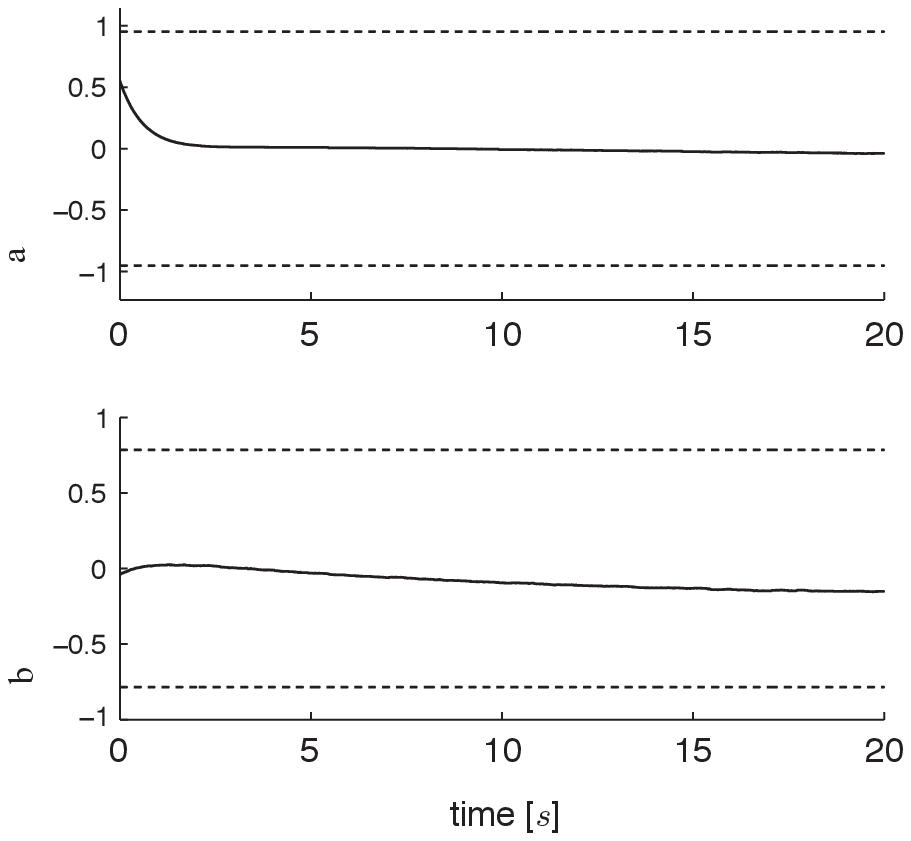}}
\end{tabular}
\end{center}
\vspace{-0.15cm}
\caption{\emph{Rejection of unknown but bounded disturbances}: (a) Trajectory of the leader and follower, and
visibility set $\mathcal{S}$; (b)~$\Delta
\subupscr{p}{L}{F}[1]$, $\subupscr{p}{L}{F}[2]$, $\subupscr{\beta}{L}{F}$ (solid)
and bounds $\pm a$, $\pm a$, $\pm b$ (dash); (c)~$\subscr{v}{F}$, $\subscr{\omega}{F}$ (solid) and bounds \mbox{$\pm
\subscr{V}{F}$, $\pm \subscr{\Omega}{F}$ (dash)}.}\label{FIG:simulaz2}
\end{figure*}
%
%
\begin{figure*}[t!]
\begin{center}
\begin{tabular}{ccc}
\!\!\!\!\!\subfigure[]{\includegraphics[width=0.71\columnwidth]{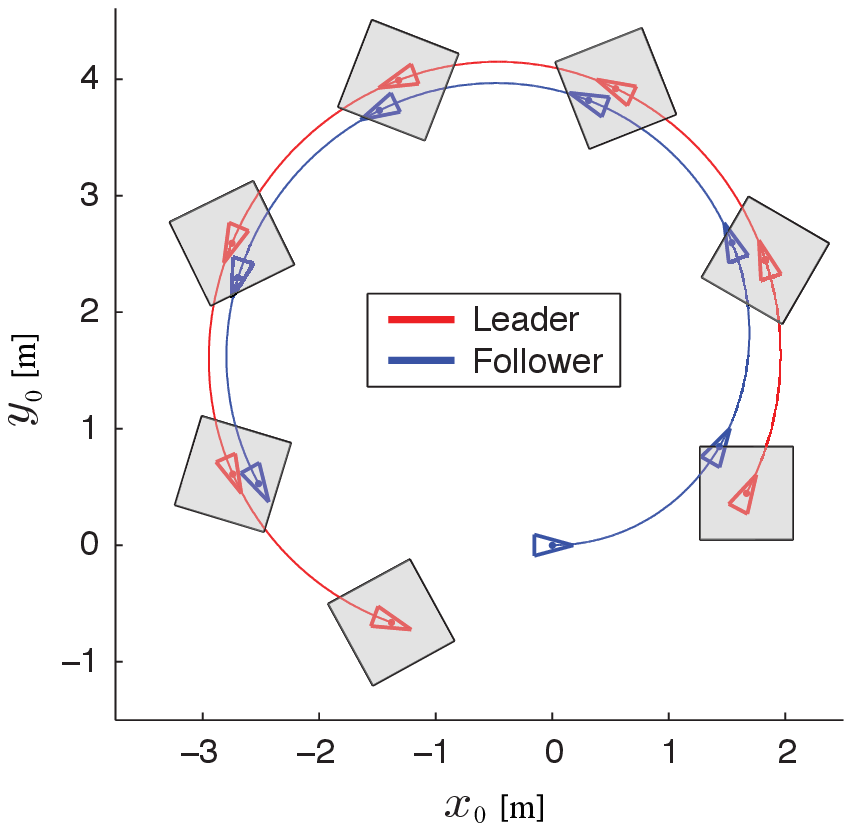}}
&
\psfrag{c}{\scriptsize{$\Delta \subupscr{p}{L}{F}[1]$\,
[m]}} \psfrag{b}{\scriptsize{$\Delta \subupscr{p}{L}{F}[2]$\, [m]}}
\psfrag{a}{\scriptsize{$\Delta \subupscr{\beta}{L}{F}$\, [rad]}}
\subfigure[]{\includegraphics[width=0.652\columnwidth]{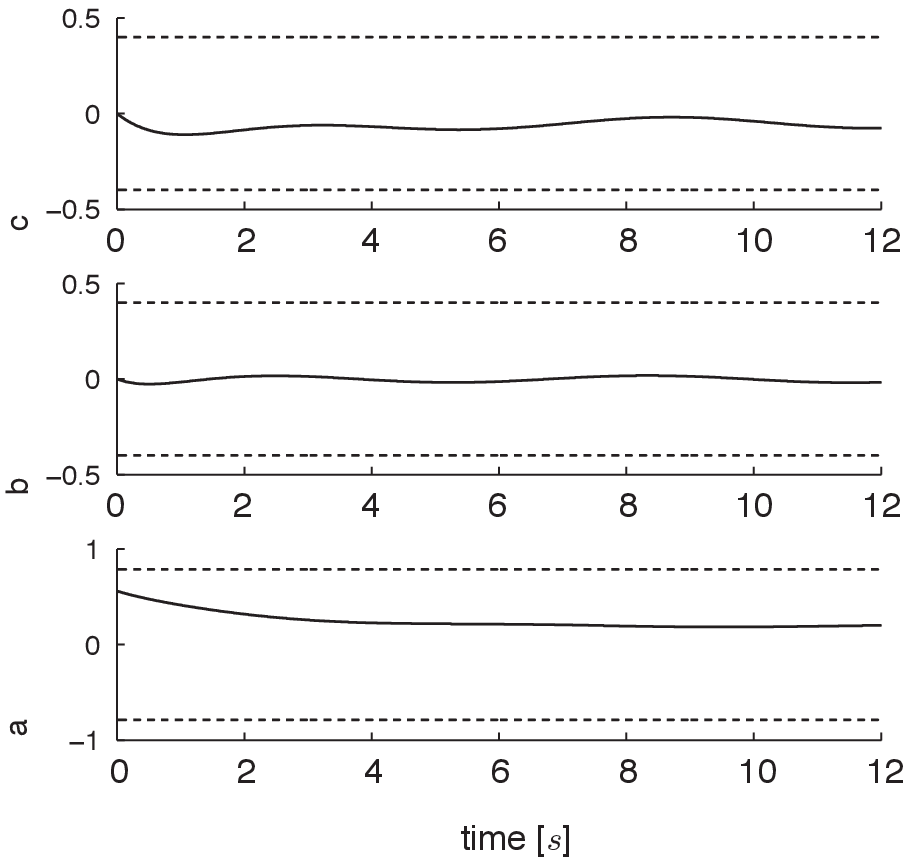}}
&
\psfrag{b}{\footnotesize{$\subscr{v}{F}$}\,
\scriptsize{[m/s]}}
\psfrag{a}{\footnotesize{$\Delta\subscr{\omega}{F}$}\,
\scriptsize{[rad/s]}}
\subfigure[]{\includegraphics[width=0.648\columnwidth]{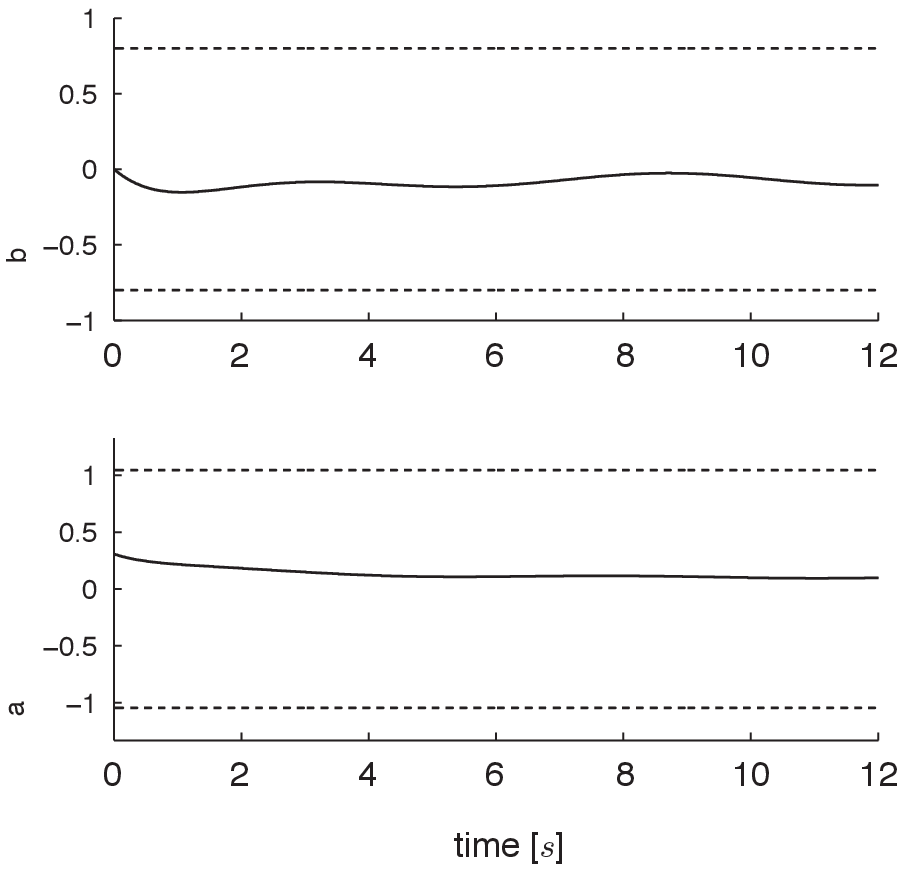}}
\end{tabular}
\end{center}
\vspace{-0.15cm}
\caption{\emph{VMP on a circle}: (a) Trajectory of the leader and follower, and
visibility set $\mathcal{S}$; (b) $\Delta\subupscr{p}{L}{F}[1]$, $\Delta\subupscr{p}{L}{F}[2]$, $\Delta\subupscr{\beta}{L}{F}$ (solid)
and bounds $\pm a$, $\pm a$, $\pm b$ (dash); (c)~$\subscr{v}{F}$, $\Delta\subscr{\omega}{F}$ (solid) and bounds \mbox{$\pm
\subscr{V}{F}$, $\pm \subscr{\Omega}{F}$ (dash)}.}\label{FIG:simulaz3}
\end{figure*}
\vspace{-0.2cm}
\section{Simulation results}\label{SEC:simul}
\vspace{-0.1cm}

Extensive simulation experiments have been performed to illustrate the
theory and assess the soundness of the~proposed~approach. 

\subsection{Basic scenario}

\indent The simulation results reported in Fig.~\ref{FIG:simulaz1} refer to the basic scenario 
studied at the beginning of Sect.~\ref{SEC:VMP}. The~leader robot moves with velocities
$\subscr{v}{L}(t) = 0.05\,\sin(t)$, $\subscr{\omega}{L}(t) = \frac{\pi}{20} \cos(0.1 t)$.
We set $\subscr{V}{L} = 0.1$~m/s,
$\subscr{\Omega}{L}=\pi/15$~rad/s, $\subscr{V}{F} = 0.9$~m/s,
$\subscr{\Omega}{F}=\pi/3$ rad/s, $a = 0.4$ m, $b=\pi/4$~rad and
$d=2$~m, according to the conditions of Theorem~\ref{Teo1} and
we~chose the gain matrix in $\mathcal{K}$ with minimum
\mbox{2-norm:} 
$$
K=\left[
\begin{array}{ccc}
1.5173 \,&\, 0 \,&\, 0\\
0 \,&\, 0.3707 \,&\, 0.4925
\end{array}
\right]\!. \vspace{-0.12cm}
$$ 
Note that since $K$ is in the interior of
$\mathcal{K}$, the asymptotic stability is assured (cf. Theorem 5.2 in~\mbox{\cite{Blanchini_JOTA91}}).
System~(\ref{EQ:nonlin_unc_param}) has been initialized with
\begin{equation}\label{Eq:init_cond}
\begin{array}{c}
\hspace{-1.3cm}(\Delta \subupscr{p}{L}{F}[1](0), \subupscr{p}{L}{F}[2](0),
\subupscr{\beta}{L}{F}(0))^T\vspace{0.16cm}\\ 
\hspace{1.3cm}= (0.3285,\, -0.1626,\, 0.1071)^T\!. \vspace{-0.12cm}
\end{array}
\end{equation}
Fig.~\ref{FIG:simulaz1}(a) reports the trajectory of robot
$\textup{L}$ and $\textup{F}$ and the visibility set
$\mathcal{S}$, (in order to have a temporal reference 
the robots are drawn every two seconds).
Fig.~\ref{FIG:simulaz1}(b) shows that $\Delta
\subupscr{p}{L}{F}[1]$, $\subupscr{p}{L}{F}[2]$, $\subupscr{\beta}{L}{F}$ (solid),
keep inside the respective bounds $\pm a$, $\pm a$, $\pm b$
(dash), as expected. Finally, Fig.~\ref{FIG:simulaz1}(c) exposes that
the control inputs $\subscr{v}{F}$, $\subscr{\omega}{F}$ (solid),
respect the corresponding bounds $\pm \subscr{V}{F}$, $\pm
\subscr{\Omega}{F}$ (dash).
%
%
\begin{figure*}[t!]
\begin{center}
\begin{tabular}{ccc}
\!\!\!\!\!\!\!\!
\psfrag{a}{\scriptsize{$\mathcal{S}_2$}}
\subfigure[]{\includegraphics[width=0.696\columnwidth]{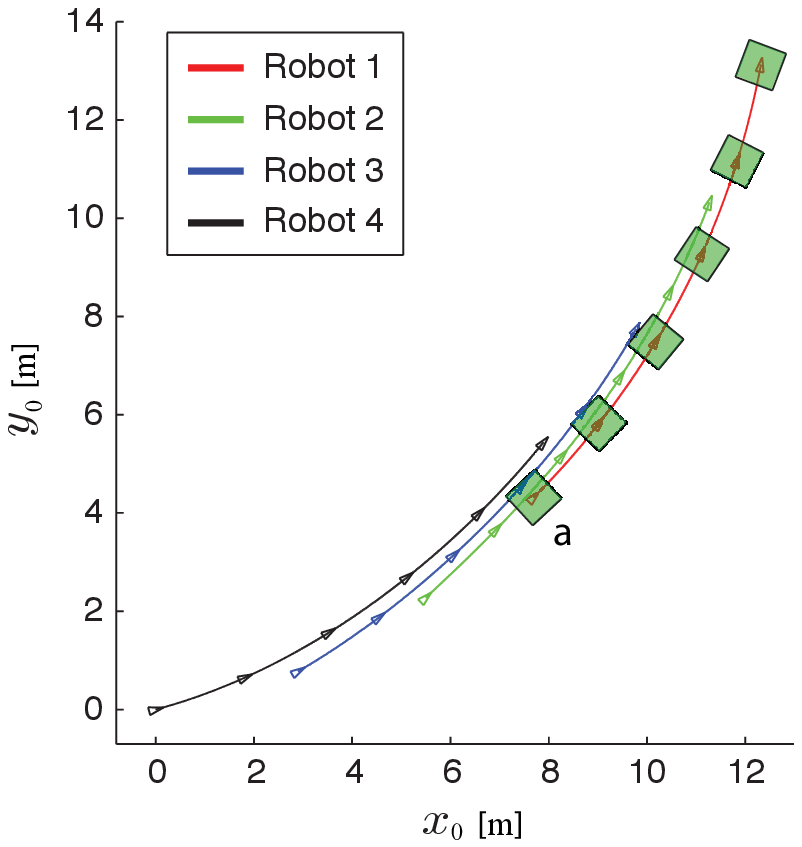}}
\!\!\!\!\!\!\!&\!\!\!\!\!
\psfrag{a}{\scriptsize{$\mathcal{S}_3$}}
\subfigure[]{\includegraphics[width=0.679\columnwidth]{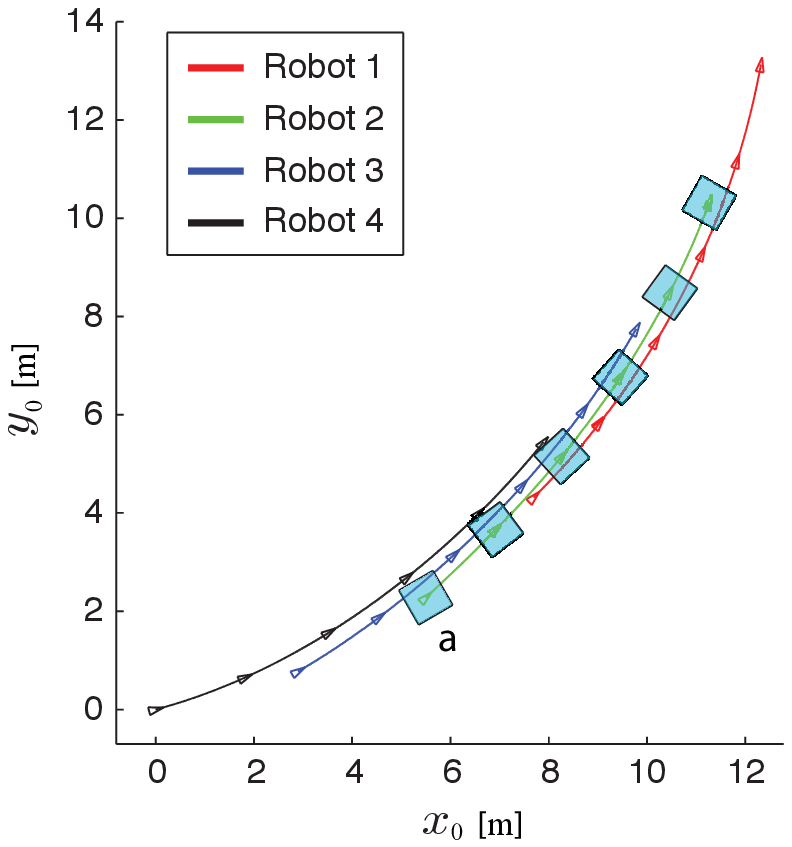}}
\!\!\!\!\!&\!\!\!\!\!
\psfrag{a}{\scriptsize{$\mathcal{S}_4$}}
\subfigure[]{\includegraphics[width=0.676\columnwidth]{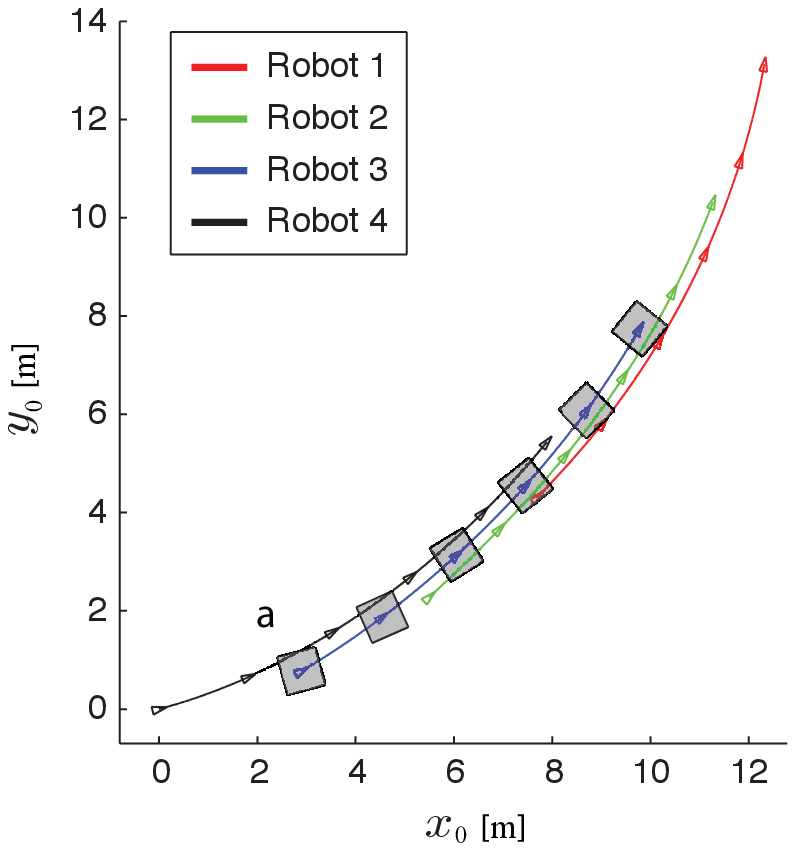}}\\
\vspace{0.2cm}
\!\!\!\!
\psfrag{c}{\scriptsize{$\Delta \subupscr{p}{1}{2}[1]$\,
[m]}} \psfrag{b}{\scriptsize{$\subupscr{p}{1}{2}[2]$\, [m]}}
\psfrag{a}{\scriptsize{$\subupscr{\beta}{1}{2}$\, [rad]}}
\subfigure[]{\includegraphics[width=0.655\columnwidth]{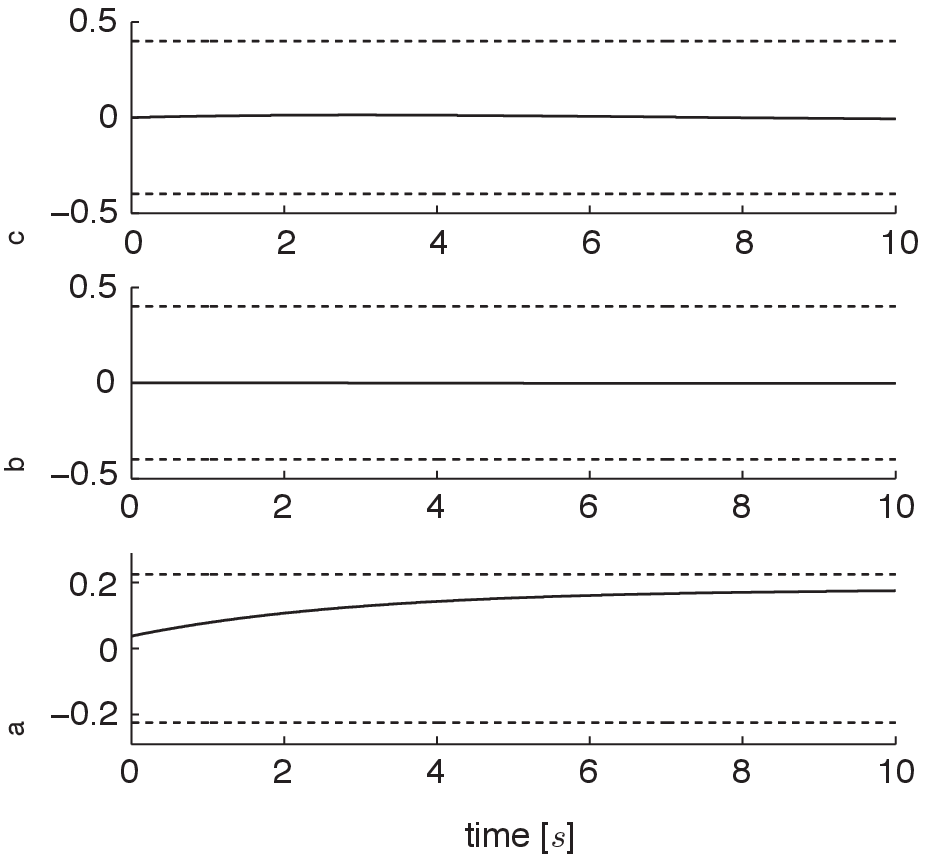}}
&
\psfrag{c}{\scriptsize{$\Delta \subupscr{p}{2}{3}[1]$\,
[m]}} \psfrag{b}{\scriptsize{$\subupscr{p}{2}{3}[2]$\, [m]}}
\psfrag{a}{\scriptsize{$\subupscr{\beta}{2}{3}$\, [rad]}}
\subfigure[]{\includegraphics[width=0.655\columnwidth]{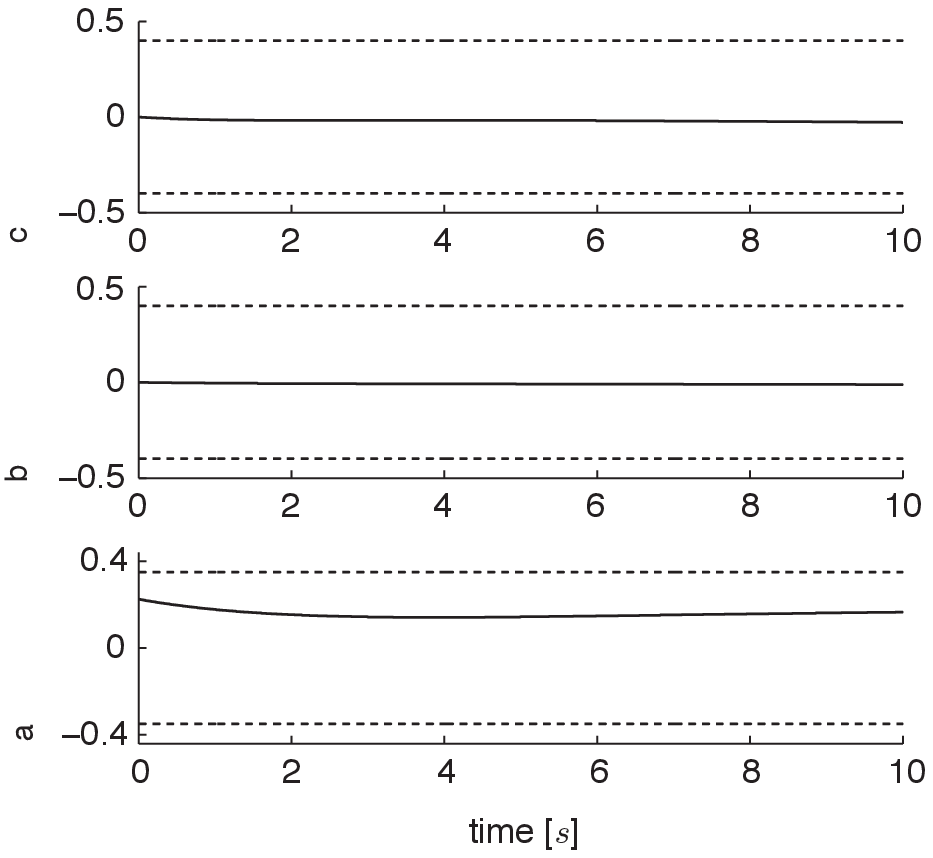}}
\!&\!\!
\psfrag{c}{\scriptsize{$\Delta \subupscr{p}{3}{4}[1]$\,
[m]}} \psfrag{b}{\scriptsize{$\subupscr{p}{3}{4}[2]$\, [m]}}
\psfrag{a}{\scriptsize{$\subupscr{\beta}{3}{4}$\, [rad]}}
\subfigure[]{\includegraphics[width=0.655\columnwidth]{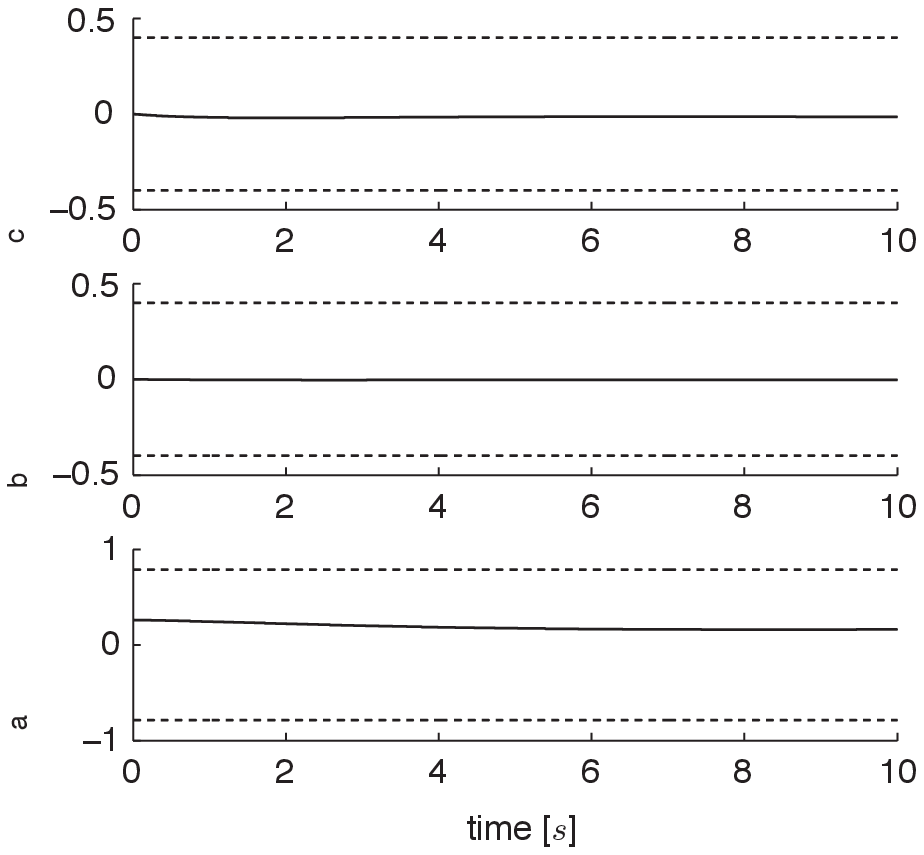}}\\
\vspace{0.1cm}
\!\!\!\!
\psfrag{b}{\footnotesize{$\subscr{v}{2}$}\,
\scriptsize{[m/s]}}
\psfrag{a}{\footnotesize{$\subscr{\omega}{2}$}\,
\scriptsize{[rad/s]}}
\subfigure[]{\includegraphics[width=0.661\columnwidth]{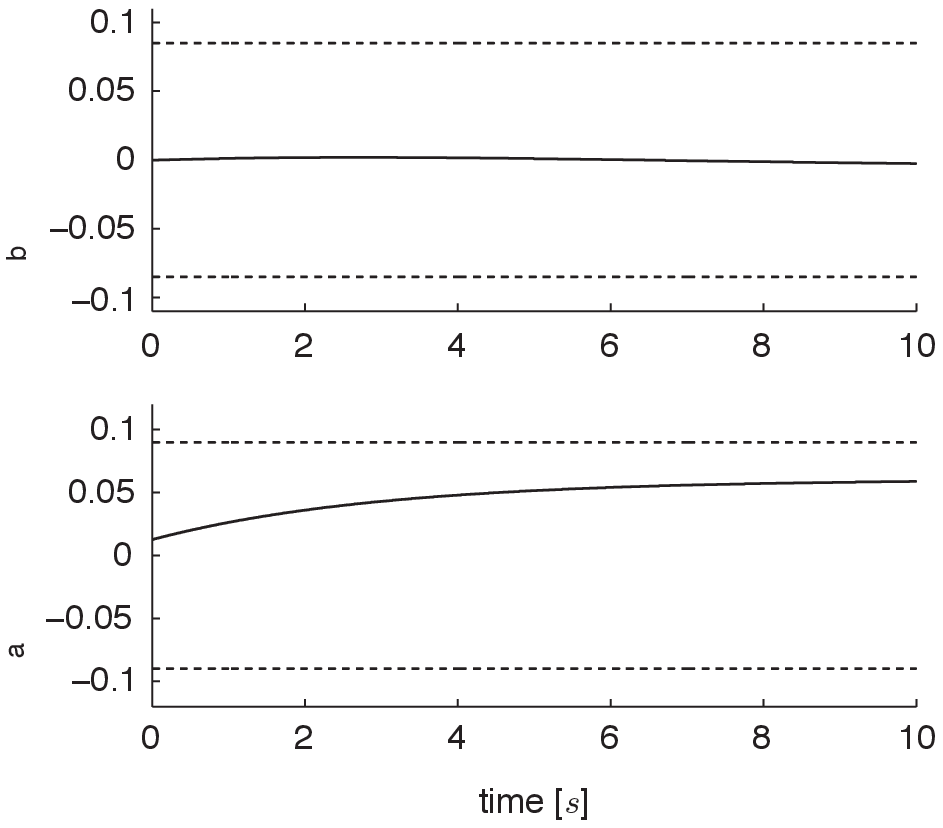}}
\!&\!
\psfrag{b}{\footnotesize{$\subscr{v}{3}$}\,
\scriptsize{[m/s]}}
\psfrag{a}{\footnotesize{$\subscr{\omega}{3}$}\,
\scriptsize{[rad/s]}}
\subfigure[]{\includegraphics[width=0.655\columnwidth]{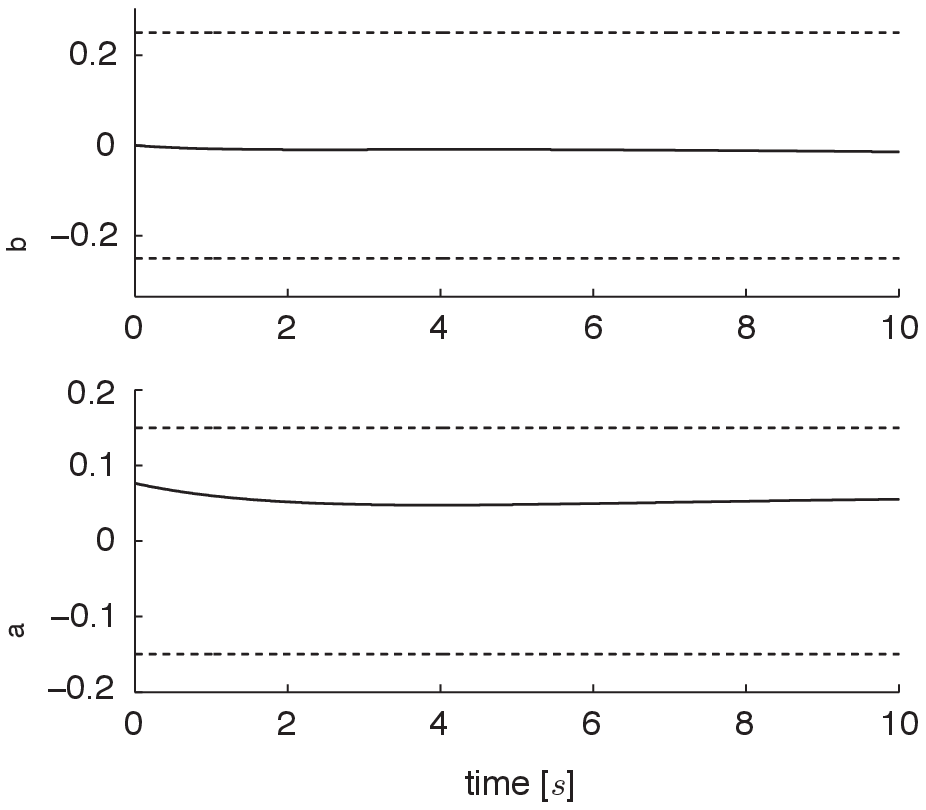}}
\!\!&\!\!\!
\psfrag{b}{\footnotesize{$\subscr{v}{4}$}\,
\scriptsize{[m/s]}}
\psfrag{a}{\footnotesize{$\subscr{\omega}{4}$}\,
\scriptsize{[rad/s]}}
\subfigure[]{\includegraphics[width=0.655\columnwidth]{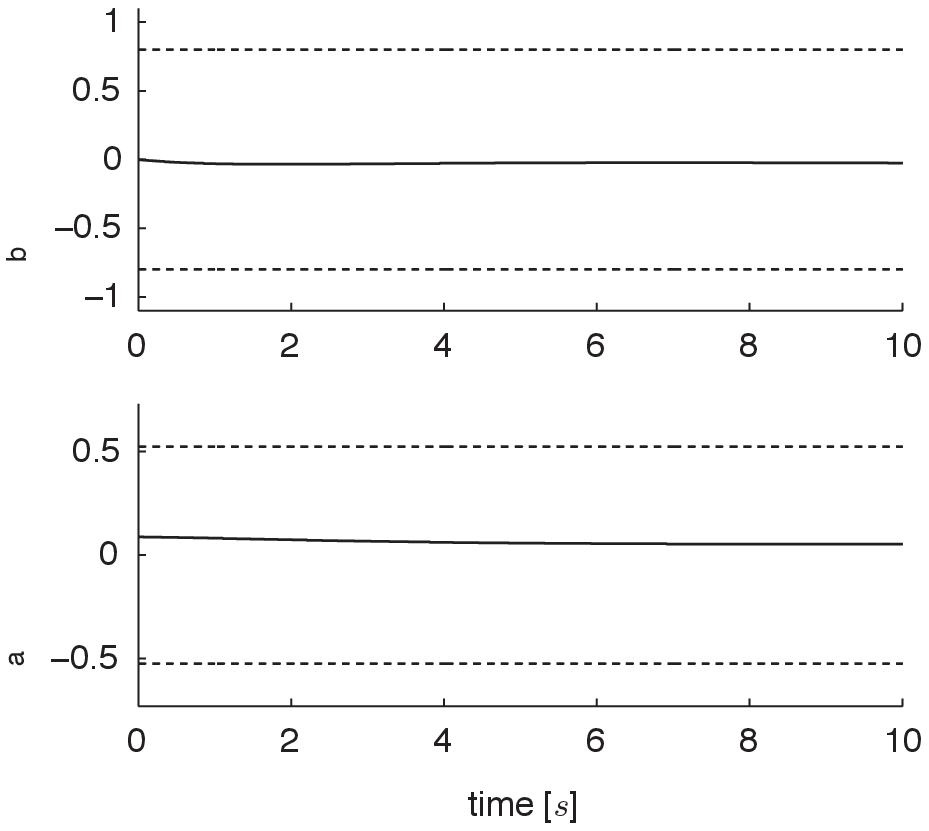}}
\end{tabular}
\end{center}
\vspace{-0.3cm}
\caption{\emph{Chain of robots}: Trajectory of the 4 robots and (a)
visibility set $\mathcal{S}_2$, (b) visibility set $\mathcal{S}_3$, (c)
visibility set $\mathcal{S}_4$; (d) $\Delta \subupscr{p}{1}{2}[1]$, 
$\subupscr{p}{1}{2}[2]$, $\subupscr{\beta}{1}{2}$ (solid)
and bounds $\pm a_2$, $\pm a_2$, $\pm b_2$ (dash); (e) $\Delta
\subupscr{p}{2}{3}[1]$, $\subupscr{p}{2}{3}[2]$, $\subupscr{\beta}{2}{3}$
and bounds $\pm a_3$, $\pm a_3$, $\pm b_3$; (f) $\Delta
\subupscr{p}{3}{4}[1]$, $\subupscr{p}{3}{4}[2]$, $\subupscr{\beta}{3}{4}$
and bounds $\pm a_4$, $\pm a_4$, $\pm b_4$; (g)
$\subscr{v}{2}$, $\subscr{\omega}{2}$ (solid) and bounds $\pm
\subscr{V}{2}$, $\pm \subscr{\Omega}{2}$ (dash); (h)
$\subscr{v}{3}$, $\subscr{\omega}{3}$ and bounds $\pm
\subscr{V}{3}$, $\pm \subscr{\Omega}{3}$;
(i) $\subscr{v}{4}$, $\subscr{\omega}{4}$ and bounds $\pm
\subscr{V}{4}$, $\pm \subscr{\Omega}{4}$.}\label{FIG:simulaz4}
\end{figure*}

\subsection{Rejection of unknown but bounded disturbances}

In the simulation results reported in Fig.~\ref{FIG:simulaz2}, the leader robot
moves with velocities
$\subscr{v}{L}(t) = 0.01$,  
$\subscr{\omega}{L}(t) = -\frac{\pi}{20} \,\sin(0.08 t)$.
Unknown but bounded disturbances $\subscr{h}{L}(t)$, $\subscr{h}{F}(t)$ 
(uniform random noises in the interval $(-0.1,\,0.1)$) act on the leader and 
the follower (recall Sect.~\ref{Subsect:LatDist}). Owing to the 
conditions of Corollary~\ref{Cor_UBB}, we chose $\subscr{V}{L} = 0.03$~m/s,
$\subscr{\Omega}{L}=\pi/18$~rad/s, $\subscr{V}{F} = 0.95$~m/s,
$\subscr{\Omega}{F}=\pi/4$ rad/s, $\subscr{H}{L} = \subscr{H}{F} = 0.12$~m/s,
$a = 0.4$ m, $b=\pi/4$~rad and $d=2$~m.
The initial condition of system (\ref{EQ:nonlin_UBB2}) is (\ref{Eq:init_cond}) and, again,
we selected the gain matrix in $\mathcal{K}$ with \mbox{minimum 2-norm:} 
$$
K=\left[
\begin{array}{ccc}
1.6735 \,&\, 0 \,&\, 0 \vspace{0.07cm}\\
0 \,&\, 0.5896 \,&\, 0.5326
\end{array}
\right]\!.
$$ 
Fig.~\ref{FIG:simulaz2}(a) reports the trajectory of
$\textup{L}$ and $\textup{F}$ and the visibility set
$\mathcal{S}$. From Figs.~\ref{FIG:simulaz2}(b) and~\ref{FIG:simulaz2}(c), we note that
despite the presence of the unknown but bounded disturbances,
$\Delta \subupscr{p}{L}{F}[1]$, $\subupscr{p}{L}{F}[2]$, $\subupscr{\beta}{L}{F}$ and
$\subscr{v}{F}$, $\subscr{\omega}{F}$
respect the relative state and control bounds.

\vspace{-0.2cm}
\subsection{VMP on a circle}
\vspace{-0.2cm}

%
In Fig.~\ref{FIG:simulaz3}, the leader robot moves with velocities
$\subscr{v}{L}(t) = 0.05\,\sin(t)$, $\subscr{\omega}{L}(t) = \pi/30$,
and the parameters $\subscr{V}{L} = 0.06$~m/s,
$\subscr{\Omega}{L}=\pi/25$~rad/s, $\subscr{V}{F} = 0.8$~m/s,
$\subscr{\Omega}{F}=\pi/3$ rad/s,
$a = 0.4$ m, $b=\pi/4$~rad, \mbox{$\rho=0.3$}~rad/s and \mbox{$\gamma=\pi/6$~rad}, 
have been chosen according to the conditions of Theorem~\ref{Theo_circ} 
(recall Sect.~\ref{Subsect:AngPar}).
\mbox{The minimum 2-norm} gain matrix in $\mathcal{K}$ is, in this case
$$
K=\left[
\begin{array}{ccc}
1.3812 \,&\, 0 \,&\, 0\\
0 \,&\, 0.6051 \,&\, 0.5508
\end{array}
\right]\!,
$$ 
and the initial condition of system (\ref{EQ_sysring}) is 
$(\Delta\subupscr{p}{L}{F}[1](0),$\, $\Delta\subupscr{p}{L}{F}[2](0),\,
\Delta\subupscr{\beta}{L}{F}(0))^T$ $= (0,\, 0,\, 0.5597)^T$.
Fig.~\ref{FIG:simulaz3}(a) reports the trajectory of
$\textup{L}$ and $\textup{F}$ and the visibility set
$\mathcal{S}$.
Figs.~\ref{FIG:simulaz3}(b) and~\ref{FIG:simulaz3}(c)
show the time history of $\Delta \subupscr{p}{L}{F}[1]$, $\Delta\subupscr{p}{L}{F}[2]$, 
$\Delta\subupscr{\beta}{L}{F}$ and $\subscr{v}{F}$, $\Delta\subscr{\omega}{F}$, and 
the corresponding~bounds.

\vspace{-0.1cm}
\subsection{Chain of robots}
\vspace{-0.1cm}

In Fig.~\ref{FIG:simulaz4} a chain of 4 robots is considered (recall \mbox{Sect.~\ref{SEC:ext_chain}}).
Robot~1 guides the formation with velocities, 
$\subscr{v}{1}(t) = 0.01$, $\subscr{\omega}{1}(t) = \pi/52$.
The following set of parameters, satisfying conditions~\mbox{(\ref{EQ:chain1})-(\ref{EQ:chain3})}, 
has been used in the simulation: 
$\subscr{V}{1} = 0.02$~m/s, $\subscr{V}{2} = 0.085$~m/s,
$\subscr{V}{3} = 0.25$~m/s, $\subscr{V}{4} = 0.8$~m/s,
$\subscr{\Omega}{1}=\pi/50$~rad/s, $\subscr{\Omega}{2}=\pi/35$~rad/s,
$\subscr{\Omega}{3}=\pi/21$~rad/s, $\subscr{\Omega}{4}=\pi/6$~rad/s,
and $a_2 = a_3 = a_4 = 0.4$~m, $b_2 = \pi/14$~rad, $b_3 = \pi/9$~rad,
$b_4 = \pi/4$~rad and $d_2 = d_3 = d_4 = 3$~m.
%
The minimum \mbox{2-norm} gain matrices in the polytopes $\mathcal{K}_2$, $\mathcal{K}_3$ 
and $\mathcal{K}_4$ of robots 2,\,3 and 4, are respectively
$$
\scriptsize
\begin{array}{ll}
\!\!K_2=\!\left[
\begin{array}{ccc}
0.2066 \!& 0 \,&\, 0\vspace{0.05cm}\\
0 \!& 0.0315 \,&\, 0.3361
\end{array}
\right]\!\!,\, & 
K_3=\!\left[
\begin{array}{ccc}
0.5087 \!& 0 \,&\, 0\vspace{0.05cm}\\
0 \!& 0.0669 \,&\, 0.3400
\end{array}
\right]\!\!,
\end{array}
$$
\vspace{-0.25cm}
$$
\scriptsize
K_4=\left[
\begin{array}{ccc}
 1.7273 & 0 & 0 \vspace{0.05cm}\\
0 & 0.2678 & 0.3348
\end{array}
\right]\!,\vspace{0.12cm}
$$ 
\normalsize
and the initial conditions for the three dynamic systems in the form (\ref{EQ:nonlin_unc_param}) are,
$(\Delta \subupscr{p}{1}{2}[1](0), \subupscr{p}{1}{2}[2](0),
\subupscr{\beta}{1}{2}(0))^T = (0,\, 0,\, 0.0374)^T$, 
$(\Delta \subupscr{p}{2}{3}[1](0), \subupscr{p}{2}{3}[2](0),
\subupscr{\beta}{2}{3}(0))^T =$ $(0,\, 0,\,$ $0.2244)^T$,
$(\Delta \subupscr{p}{3}{4}[1](0), \subupscr{p}{3}{4}[2](0),
\subupscr{\beta}{3}{4}(0))^T = (0,\, 0,\, 0.2618)^T$,
where $p_{k}^{k+1}$ denotes the position of robot $k$ with respect to the reference 
frame attached to robot $k+1$ and $\beta_{k}^{k+1} \triangleq {\theta_{k} - \theta_{k+1}}$, $k \in \{1,2,3\}$.
Figs.~\ref{FIG:simulaz4}(a)-(c) report the trajectory of the 4~robots
and the visibility sets $\mathcal{S}_2$ (green), $\mathcal{S}_3$ (cyan) 
and $\mathcal{S}_4$ (gray), respectively.
Figs.~\ref{FIG:simulaz4}(d)-(f) and~\ref{FIG:simulaz4}(g)-(i)
show the time history of
$\Delta p_{k}^{k+1}[1]$, $p_{k}^{k+1}[2]$, $\beta_{k}^{k+1}$, $k \in \{1,2,3\}$ and 
$v_{k}$, $\omega_{k}$, $k \,{\in}\, \{2,3,4\}$, and the relative state and control bounds.

\vspace{-0.1cm}
\section{Conclusions and future work}\label{SEC:ConcFut}
\vspace{-0.15cm}

The paper proposes an original solution to the \emph{visibility
maintenance problem} (VMP) for a leader-follower pair of Dubins-like
vehicles with input constraints. 
\mbox{By interpreting} the nonlinear model describing the relative dynamics
of the robots as a linear system with parameter uncertainty,
the~VMP is reformulated as a \emph{linear constrained regulation
problem with additive \mbox{disturbances}} (DLCRP). General conditions
for the \mbox{positive} $\mathcal{D}$-invariance of linear
uncertain systems with parametric disturbance matrix are derived
and used to study the feasibility of the VMP when box bounds on
the state, input and disturbance are considered.
The~proposed design procedure can be easily adapted to provide the 
control with unknown but bounded disturbances rejection capabilities. Conditions for the 
solution of the VMP when robots' desired displacement is defined through angular 
parameters are also presented, and the extension to 
chains of $n$ robots is discussed.\\
A drawback of the approach proposed in this paper is that it requires 
a great amount of computational work off-line, because the feedback 
matrix $K$ is obtained as a solution of a large set of inequalities. 
In addition all the vertices of the visibility set $\mathcal{S}$ are required. 
In~this respect, the solution to the LCRP proposed in 
\cite{VassilakiBi_SCL89} appears to be preferable to the one in~\cite{Blanchini_TAC90}, 
even though neither disturbances nor model parametric uncertainty 
are considered therein.\\ 
Future research lines include the extension of our results to robotic
networks with arbitrary topologies and the application of the 
proposed approach to the study of consensus, rendezvous and 
coverage problems in the presence of visibility constraints. 
The use of polar coordinates 
to describe conic-like visibility sets $\mathcal{S}$, is also 
a subject of on-going research. 

\vspace{-0.17cm}
\section*{Appendix A: The Fourier-Motzkin elimination method} 

The Fourier-Motzkin elimination (FME), a generalization of Gauss elimination, 
is a computational method for solving a system
$$
A\, x \leq b,\;\; A \in \mathbb{R}^{m \times n},\;\; b \in \mathbb{R}^{m},
$$ 
of $m$ linear inequalities in $n$ variables~\cite{Motzkin_51,Book_Schrijver}.
The key idea of the FME method is to eliminate one variable of the
system $A\,x \leq b$ at each iteration and rewrite the resulting
equations accordingly. 
Even though the number of variables decreases at each step, the number of inequalities in the
remaining variables grows exponentially fast: in fact, at~iteration $j$ the 
number of inequalities to be evaluated is at most $\lfloor
\frac{m}{2}\rfloor^{2^j}$. 
Because of its double-exponential computational complexity, the FME method can be applied efficiently only to
problems with a small number of inequalities and it is not competitive with standard 
LP solvers
\mbox{However},~differently from this numerical approach, the method 
can handle \emph{symbolic inequalities}. 
We illustrate this idea with a simple example. 
Consider the following set of \mbox{symbolic inequalities}
\begin{align}
a_{11} \,x_1 + a_{12} \,x_2 &\,\leq\, b_{1},\label{Eq_sys_example1}\\
- a_{21} \,x_1 - a_{22} \,x_2 &\,\leq\, b_{2},\label{Eq_sys_example2}\\
a_{32} \,x_2 &\,\leq\, b_3,\label{Eq_sys_example3}
\end{align}
where $a_{11}$, $a_{12}$, $a_{21}$, $a_{22}$, $a_{32}$, $b_1$, $b_2$, $b_3 \in \mathbb{R}_{>0}$ are unknown parameters.
We wish to determine under which conditions on these parameters, system (\ref{Eq_sys_example1})-(\ref{Eq_sys_example3})
admits solutions. The first step is to eliminate the variable
$x_1$. To this end, solving (\ref{Eq_sys_example1})-(\ref{Eq_sys_example2}) in $x_1$, we get: 
$$
\begin{array}{c}
\displaystyle \!\!\!\!\!x_1 \leq \frac{b_1 - a_{12}\,x_2}{a_{11}},\quad
\displaystyle x_1 \geq \frac{-b_2 - a_{22}\,x_2}{a_{21}}. 
\end{array}
$$
Joining the two inequalities, after a few computations, we obtain the following condition on the variable $x_2$ 
(\mbox{assuming} that $a_{11}\,a_{22} - a_{12}\,a_{21} \neq 0$)
\begin{equation}\label{Eq_sys_example4}
x_2 \geq \frac{-a_{11}\,b_2 - a_{21}\,b_1}{a_{11}\,a_{22} - a_{12}\,a_{21}}.
\end{equation}
If we finally combine inequalities (\ref{Eq_sys_example3}) and (\ref{Eq_sys_example4}), we end up with the 
sought solvability condition of system (\ref{Eq_sys_example1})-(\ref{Eq_sys_example3}), in terms of the symbolic 
parameters $a_{11}$, $a_{12}$, $a_{21}$, $a_{22}$, $a_{32}$, $b_1$, $b_2$, $b_3$:
$$
\frac{b_3}{a_{32}} \geq \frac{-a_{11}\,b_2 - a_{21}\,b_1}{a_{11}\,a_{22} - a_{12}\,a_{21}}. \vspace{-0.4cm}
$$

\section*{Appendix B: Application of the Fourier-Motzkin elimination to the \mbox{inequalities}
  (\ref{Eq:All_cond1})-(\ref{Eq:All_cond2})}
\vspace{-0.2cm}

Our first step to solve the system of inequalities
\mbox{(\ref{Eq:All_cond1})-(\ref{Eq:All_cond2})}, 
is to eliminate variable $k_{11}$.
Solving in $k_{11}$ (under the assumption of $d > a$), we obtain the following set of conditions:
\vspace{-0.25cm}
$$
\begin{array}{l}
\displaystyle k_{11} \geq q_4 \,k_{22} + \frac{b}{a} \,q_4 \,k_{23} + \frac{b}{a} \,q_2 + \frac{\subscr{V}{L}}{a},\vspace{0.2cm}\\
\displaystyle k_{11} \geq q_4 \,k_{22} - \frac{b}{a} \,q_4 \,k_{23} - \frac{b}{a} \,q_2 + \frac{\subscr{V}{L}}{a},\vspace{0.2cm}\\
\displaystyle k_{11} \geq -q_4 \,k_{22} + \frac{b}{a} \,q_4 \,k_{23} + \frac{b}{a} \,q_2 + \frac{\subscr{V}{L}}{a},\vspace{0.2cm}\\
\displaystyle k_{11} \geq -q_4 \,k_{22} - \frac{b}{a} \,q_4 \,k_{23} - \frac{b}{a} \,q_2 + \frac{\subscr{V}{L}}{a},\vspace{0.2cm}\\
\displaystyle k_{11} \geq -\frac{\subscr{V}{F}}{a},\;\; k_{11} \leq \frac{\subscr{V}{F}}{a}.
\end{array}
$$
Combining these inequalities, we get: 
$$
\begin{array}{l}
\displaystyle q_4 \,k_{22} \leq \frac{\subscr{V}{F}}{a} - \frac{b}{a} \,q_4 \,k_{23} - \frac{b}{a} \,q_2 - \frac{\subscr{V}{L}}{a},\vspace{0.3cm}\\
\displaystyle q_4 \,k_{22} \leq \frac{\subscr{V}{F}}{a} + \frac{b}{a} \,q_4 \,k_{23} + \frac{b}{a} \,q_2 - \frac{\subscr{V}{L}}{a},\vspace{0.3cm}\\
\displaystyle q_4 \,k_{22} \geq -\frac{\subscr{V}{F}}{a} + \frac{b}{a} \,q_4 \,k_{23} + \frac{b}{a} \,q_2 + \frac{\subscr{V}{L}}{a},\vspace{0.3cm}\\
\displaystyle q_4 \,k_{22} \geq -\frac{\subscr{V}{F}}{a} - \frac{b}{a} \,q_4 \,k_{23} - \frac{b}{a} \,q_2 + \frac{\subscr{V}{L}}{a}.
\end{array}
$$
In order to eliminate the second variable, $k_{22}$, we should consider three cases, according to the sign of $q_4$.
If we assume that $q_4 > 0$, we then obtain the following set of
\mbox{inequalities}:
\vspace{-0.25cm}
\begin{align}
k_{22} &\geq - \frac{b}{a} \,k_{23} + \frac{b\,(1+q_1) + \subscr{V}{L} \,\sin b}{a(d+q_3)},\label{Eq.AppB1}\\
k_{22} &\geq  \frac{b}{a} \,k_{23} + \frac{-b\,(1+q_1) + \subscr{V}{L} \,\sin b}{a(d+q_3)},\\
k_{22} &\geq  -\frac{b}{a} \,k_{23} + \frac{\subscr{\Omega}{L}}{a},\\
k_{22} &\geq  \frac{b}{a} \,k_{23} + \frac{-\subscr{V}{F} + b \,q_2 + \subscr{V}{L}}{a q_4},\\
k_{22} &\geq  -\frac{b}{a} \,k_{23} + \frac{-\subscr{V}{F} - b \,q_2 + \subscr{V}{L}}{a q_4},\\
k_{22} &\geq  \frac{b}{a} \,k_{23} - \frac{\subscr{\Omega}{F}}{a},\;\;
k_{22} \geq  -\frac{b}{a} \,k_{23} - \frac{\subscr{\Omega}{F}}{a},\label{Eq.AppB2}\\
k_{22} &\leq  \frac{b}{a} \,k_{23} - \frac{\subscr{\Omega}{L}}{a},\label{Eq.AppB3}\\
k_{22} &\leq  -\frac{b}{a} \,k_{23} + \frac{\subscr{V}{F} - b\, q_2 - \subscr{V}{L}}{a q_4},\\
k_{22} &\leq  \frac{b}{a} \,k_{23} + \frac{\subscr{V}{F} + b\, q_2 - \subscr{V}{L}}{a q_4},\\
k_{22} &\leq  -\frac{b}{a} \,k_{23} + \frac{\subscr{\Omega}{F}}{a},\;\;
k_{22} \leq  \frac{b}{a} \,k_{23} + \frac{\subscr{\Omega}{F}}{a}.\label{Eq.AppB4}
\end{align}
Combining conditions (\ref{Eq.AppB1})-(\ref{Eq.AppB2}) and (\ref{Eq.AppB3})-(\ref{Eq.AppB4}), 
we end up with a total set of 35 inequalities, of whom only 4 are \mbox{non-trivial}:
\vspace{-0.1cm}
$$
\begin{array}{l}
\displaystyle\subscr{\Omega}{L}  \leq \frac{b\,(1+q_1) -
\subscr{V}{L}\, \sin b}{d+q_3},\;\;  
\displaystyle\subscr{\Omega}{F} \geq \frac{b\,(1+q_1) + \subscr{V}{L}\, \sin b}{d+q_3},\vspace{0.35cm}\\
\displaystyle\subscr{V}{F} \geq \subscr{V}{L}\Big(1 + \frac{q_4 \sin b}{d+q_3}\Big) \!+ 
b \Big(q_2 + \frac{q_4 (1+q_1)}{d+q_3}\Big),\vspace{0.35cm}\\
\displaystyle\subscr{V}{F} \geq \subscr{V}{L}\Big(1 + \frac{q_4 \sin
  b}{d+q_3}\Big)\!- b \Big(q_2 + \frac{q_4 (1+q_1)}{d+q_3}\Big).
\end{array}
$$
If we now assume that $q_4 = 0$, we obtain the unique condition:
$\subscr{V}{F} \geq \subscr{V}{L} + b\,q_2$. Finally, if we assume
that~\mbox{$q_4 < 0$}, we come up with the following inequalities
\vspace{-0.1cm}
\begin{align*}
k_{22} &\geq - \frac{b}{a} \,k_{23} + \frac{b\,(1+q_1) + \subscr{V}{L} \,\sin b}{a(d+q_3)},\\
k_{22} &\geq  \frac{b}{a} \,k_{23} + \frac{-b\,(1+q_1) + \subscr{V}{L} \,\sin b}{a(d+q_3)},\\
k_{22} &\geq  -\frac{b}{a} \,k_{23} + \frac{\subscr{\Omega}{L}}{a},\;\;\;\;
k_{22} \geq  -\frac{b}{a} \,k_{23} + \frac{\subscr{V}{F} - b\, q_2 - \subscr{V}{L}}{a q_4},\\
k_{22} &\geq  \frac{b}{a} \,k_{23} + \frac{\subscr{V}{F} + b\, q_2 - \subscr{V}{L}}{a q_4},\;\;\;\;
k_{22} \geq  \frac{b}{a} \,k_{23} - \frac{\subscr{\Omega}{F}}{a},\\
k_{22} &\geq  -\frac{b}{a} \,k_{23} - \frac{\subscr{\Omega}{F}}{a},\;\;\;\;
k_{22} \leq  \frac{b}{a} \,k_{23} - \frac{\subscr{\Omega}{L}}{a},\\
k_{22} &\leq  \frac{b}{a} \,k_{23} + \frac{-\subscr{V}{F} + b \,q_2 + \subscr{V}{L}}{a q_4},\\
k_{22} &\leq  -\frac{b}{a} \,k_{23} + \frac{-\subscr{V}{F} - b \,q_2 + \subscr{V}{L}}{a q_4},\\
k_{22} &\leq  -\frac{b}{a} \,k_{23} + \frac{\subscr{\Omega}{F}}{a},\;\;\;\;
k_{22} \leq  \frac{b}{a} \,k_{23} + \frac{\subscr{\Omega}{F}}{a},
\end{align*}
from which we deduce the following two new conditions:
\vspace{-0.05cm}
$$
\begin{array}{c}
\displaystyle\subscr{V}{F} \geq \subscr{V}{L}\Big(1 - \frac{q_4 \,\sin
  b}{d+q_3}\Big) + b \Big(q_2 + \frac{q_4 \,(1+q_1)}{d+q_3}\Big),\vspace{0.35cm}\\
\displaystyle\subscr{V}{F} \geq \subscr{V}{L}\Big(1 - \frac{q_4 \,\sin
  b}{d+q_3}\Big) - b \Big(q_2 + \frac{q_4 \,(1+q_1)}{d+q_3}\Big).
\end{array}
$$
\vspace{-0.8cm}
\bibliographystyle{plain} 
\bibliography{biblio_connect}
\end{document}